\documentclass[aps,reprint,
 superscriptaddress, onecolumn
]{revtex4-2}
\usepackage{graphicx}
\usepackage{newtxtext}
\usepackage{newtxmath}
\usepackage{bm}
\usepackage{natbib}
\usepackage{hyperref}
\hypersetup{
    colorlinks = true,
    urlcolor   = blue,
    citecolor  = black,
}
\usepackage{makecell}
\usepackage{subcaption}
\usepackage{array}
\newcommand{\PreserveBackslash}[1]{\let\temp=\\#1\let\\=\temp}
\newcolumntype{C}[1]{>{\PreserveBackslash\centering}p{#1}}
\newcolumntype{R}[1]{>{\PreserveBackslash\raggedleft}p{#1}}
\newcolumntype{L}[1]{>{\PreserveBackslash\raggedright}p{#1}}

\newcommand{\RomanNumeralCaps}[1]

\begin{document}

\title{Autophoresis of a Janus particle near a planar wall: a lubrication limit}

\author{Tachin Ruangkriengsin}
\email{tachin@princeton.edu}
\affiliation{Program in Applied and Computational Mathematics, Princeton University, New Jersey 08544, USA}

\author{G\"{u}nther Turk}
\affiliation{Princeton Materials Institute, Princeton University, New Jersey 08544, USA}

\author{Howard A. Stone}
\email{hastone@princeton.edu}
\affiliation{Department of Mechanical and Aerospace Engineering, Princeton University, New Jersey 08544, USA}

\begin{abstract}
We study the self-diffusiophoresis of a spherical chemically active particle near a planar, impermeable wall, with a focus on the influence of particle orientation on propulsion. We analyze a Janus particle with asymmetric surface chemical activity, consisting of a small inert region within a catalytically active cap. While numerical simulations have been used to study such particles, they encounter difficulties  resolving the flow and transport in the near-wall regime due to geometric confinement and steep solute concentration gradients. We address this limitation through an asymptotic analysis in the lubrication limit, where the gap between the particle and the wall is narrow. In particular, we consider the distinguished limit in which the inert region is asymptotically comparable in size to the lubrication region. We analyze an axisymmetric configuration in which the inert face is oriented parallel to the wall and extend the analysis to slightly tilted orientations. We find that the cap size determines whether a tilted particle rotates back toward the axisymmetric state or continues to reorient, thereby characterizing its rotational stability in the near-contact regime.

\end{abstract}
\maketitle
\section{Introduction}
Phoretic motion refers to the transport of colloidal particles driven by gradients in physicochemical fields, such as electric potential, solute concentration, or temperature \citep{anderson1989colloid}. These fields interact with the particle within a narrow interfacial layer, generating an effective slip velocity at the particle surface. Neglecting inertial effects, the force-free and torque-free conditions on a freely suspended particle then determine its rigid-body motion.

When the gradients driving phoretic motion are generated by surface reactions at the particle boundary, the resulting mechanism is known as autophoresis, or self-diffusiophoresis, when the relevant field is the chemical concentration. This mechanism was first proposed theoretically by~\citet{golestanian2007designing}, who demonstrated that asymmetric production or consumption of solute at the particle surface can lead to self propulsion. Within this framework, the active portions of the surface are modeled by prescribing a fixed solute flux density, while the inert portions satisfy the familiar no-flux condition. More recent studies have extended this classical model to incorporate the effects of solute advection and reaction on phoretic propulsion ~\citep{cordova2008osmotic, julicher2009generic,sabass2012dynamics,michelin2014phoretic}. In the present work, however, we return to the simplest model introduced by ~\citet{golestanian2007designing} and focus on the interplay between the anisotropic surface chemistry and confinement.

In an unbounded fluid, sustained autophoretic motion requires an asymmetric distribution of surface activity. This requirement has motivated extensive efforts to design and fabricate artificial swimmers with deliberately patterned surface chemistry~\citep{ebbens2010pursuit}. An experimental realization of such surface heterogeneity is provided by half-coated colloids (Janus particles), in which one hemisphere is covered with catalytic materials such as platinum. When hydrogen peroxide is added to the surrounding solution, the platinum-coated surface catalyzes its decomposition, generating chemical gradients that drive the particle’s phoretic motion. Experimental studies have demonstrated that these particles exhibit persistent motion and enhanced diffusivity, in agreement with proposed diffusiophoretic mechanisms ~\citep{paxton2004catalytic, howse2007self, jiang2010active, ebbens2011direct, ebbens2012size}.

In many practical settings, Janus particles operate in confined environments, where nearby boundaries modify the translational and rotational symmetries of their motion. This sensitivity to confinement can be exploited to steer and direct Janus particles through wall-induced hydrodynamic and diffusive interactions. In particular, chemically patterned or topographically structured boundaries have been shown to guide particle trajectories and enable controlled transport and sorting~\citep{volpe2011microswimmers,kreuter2013transport,das2015boundaries,brown2016swimming,uspal2016guiding,simmchen2016topographical}. Near-wall dynamics can be highly nontrivial, with particles exhibiting behaviors such as skating, hovering, or reflection from the boundary, which can complicate the robust design of both particle properties and boundary geometries~\citep{uspal2015self,ibrahim2016walls,mozaffari2016self}.

A variety of theoretical approaches have been developed to analyze the behavior of Janus particles near solid boundaries. These include boundary-element methods~\citep{uspal2015self,simmchen2016topographical,bayati2019dynamics,das2020floor,uspal2019active}, multipole expansion techniques~\citep{ibrahim2015dynamics,ibrahim2016walls}, eigenfunction expansions in bispherical coordinates~\citep{mozaffari2016self}, and Galerkin-based formulations~\citep{turk_singh_stone_2025}. Due to the presence of steep solute concentration gradients, these approaches
face challenges in resolving Janus particle dynamics in the extreme near-wall regime, where the particle–wall separation is small compared to the particle size. For example, \citet{uspal2015self} restrict their analysis to particle–wall separations exceeding $1.1$ particle radii, as boundary-element methods become computationally expensive at smaller gaps. Similarly, \citet{ibrahim2016walls} and \citet{turk_singh_stone_2025} introduce short-range repulsive potentials to prevent particle–wall contact when employing multipole expansion and Galerkin-based formulations, respectively, 
a remedy which may degrade the method accuracy
in the near-contact regime. Eigenfunction expansions in bispherical coordinates~\citep{mozaffari2016self} provide one of the few frameworks capable of resolving particle–wall interactions down to vanishing separations without artificial regularization. This technique, however, is restricted to simple geometries. In any event, the reliance of the semi-analytical methods on infinite series provides limited physical intuition.

In this work, we adopt an asymptotic approach to study the dynamics of a Janus particle near a wall in regimes that are difficult to access numerically. For typical Janus particle systems with colloid sizes of 1--5 microns, the regime considered here corresponds to particle--wall separations of a few hundred nanometers or less, a range that is experimentally accessible~\citep{simmchen2016topographical, ketzetzi2020diffusion, ketzetzi2020slip}. Of particular relevance are the asymptotic analyses of \citet{yariv2016wall,yariv2017boundary}, who examined the autophoretic motion of an isotropically active particle in the vicinity of a solid wall. 
While \citet{yariv2016wall,yariv2017boundary} analyzed both remote and near-wall configurations for isotropically active particles, we restrict attention here to the near-wall regime for a Janus particle, a setting that has been studied in less detail. In this regime, the leading-order particle motion is governed by solute and flow transport within the narrow lubrication gap. \citet{yariv2017boundary} showed that the near-wall behavior of an isotropically active particle coincides with bispherical-coordinate solutions for a Janus particle~\citep{mozaffari2016self} when the lubrication region lies entirely within the active cap. Here, we extend this analysis to a Janus geometry with an inert patch of size comparable to the lateral extent of the gap, thereby capturing the combined influence of the active cap and inert face on autophoretic motion near a wall.

The remaining sections of this paper are organized as follows. In Sec.~\ref{Formulation}, we formulate the problem and provide the scalings of the variables and the governing equations. In Sec.~\ref{Axi_section}, we analyze an axisymmetric configuration in which the inert face of the particle is oriented parallel to the wall. We resolve the solute concentration and flow fields within the narrow gap and obtain explicit expressions for the hydrodynamic force and the phoretic velocity normal to the wall. In Sec.~\ref{Tilted_section}, we extend the analysis to situations in which the inert face is slightly tilted, leading to fully three-dimensional solute concentration and flow fields. We quantify the effect of the tilt on the phoretic velocity parallel to the wall and on the rotational velocity about the tilt axis. We conclude with a discussion of the results in Sec.~\ref{conclusion}.

\section{Problem formulation}\label{Formulation}
\begin{figure}
    \centering
    \includegraphics[width=0.7\columnwidth]{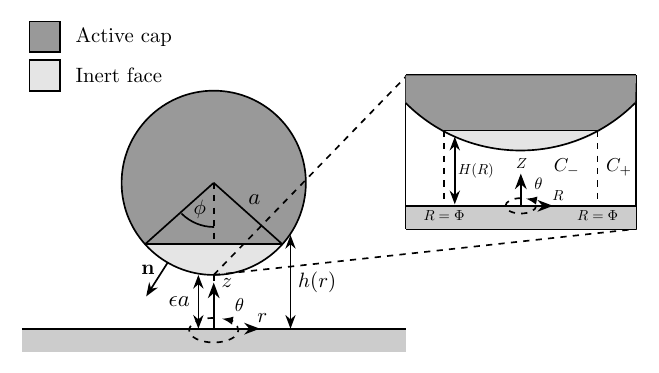}
    \caption{
       Schematic illustration of a spherical Janus particle with radius $a$ near a planar wall. 
    }
    \label{fig:schematic}
\end{figure}

We investigate the motion of a spherical Janus particle of radius $a$ near a planar wall, immersed in a solution with viscosity $\mu$. 
{ We employ cylindrical coordinates $(r,\theta,z)$ with corresponding unit vectors ($\textbf{e}_r, \textbf{e}_\theta, \textbf{e}_z$), placing the origin on the wall directly beneath the closest point on the particle surface (see Fig.~\ref{fig:schematic}). In an instantaneous configuration, the axisymmetric gap height between the particle surface and the wall at radial distance $r$ is denoted by $h(r)$, with the minimum separation occurring at $r=0$, where $h(r = 0)= h_0 = \epsilon a$.}

The particle’s surface activity establishes a solute concentration distribution in the fluid, with an associated far-field concentration denoted by $c_{\infty}.$ In the bulk, we define $c$ to be the excess solute concentration relative to $c_\infty$, such that $c$ decays in the far field. Neglecting advection, the excess solute concentration satisfies the Laplace equation,
\begin{equation}\label{laplace_dimensional}
\nabla^2 c = 0.
\end{equation}
We focus on an autophoretic particle with a catalytic active cap that produces a solute flux and an inert face that does not. We consider the simplest model where the solute flux is assumed to be uniform along the active surface, 
\begin{equation}
    -D\textbf{n} \cdot \nabla c(\textbf{x})=
    \begin{cases}
        \alpha,\quad \textbf{x} \in\text{ active cap,}\\
        0,\quad \,\textbf{x}\in\text{ inert face,}
    \end{cases}
    \label{activity}
\end{equation}
where $D$ is the solute diffusivity, $\mathbf{n}$ is the unit normal vector directed into the fluid, and $\alpha$ is the constant surface activity, with $\alpha>0$ corresponding to solute production. It is convenient to define an angle $\phi$ to be the angular span of the inert surface, see Fig. \ref{fig:schematic}. In experimentally synthesized Janus particles, the coating edge is rarely perfectly sharp and may feature a transition gradient or roughness. Since typical coating thicknesses are much smaller than the cap scale considered below~\citep{campbell2013gravitaxis}, such imperfections are not expected to alter the leading-order particle dynamics.

We assume that the Reynolds number is small, so that inertial effects are negligible. The fluid velocity field {$\textbf{u} = u_r \textbf{e}_r + u_\theta \textbf{e}_\theta + u_z \textbf{e}_z$} and pressure $p$ then satisfy the continuity equation and the Stokes equations,
\begin{equation}\label{Stokes_equations}
    \nabla \cdot \textbf{u} = 0 \quad \text{and} \quad \nabla p = \mu \nabla^2 \textbf{u}.
\end{equation}
All equations are written in the laboratory frame, in which the wall is stationary.

Interactions between the particle and the solute lead to  diffusiophoresis, in which solute concentration gradients drive an effective slip flow at the boundary of the particle,
\begin{equation}\label{phoretic_slip}
\textbf{u}_{\text{slip}}(\textbf{x}) = { \beta}\nabla_s c(\textbf{x}).
\end{equation}
Here, $\nabla_s = (\mathbf{I} - \mathbf{n}\mathbf{n}) \cdot \nabla$ denotes the surface gradient operator and {$\beta$} is the local phoretic mobility, which is assumed uniform. 
The slip-velocity model assumes that the interaction layer, of thickness $\lambda$, is thin relative to the gap scale, requiring $\lambda/a \ll \epsilon$. For neutral diffusiophoresis with short-ranged interactions, \citet{anderson1989colloid} describes this layer as molecular in thickness. Taking $\lambda$ to be of order one angstrom gives $\lambda/a\approx 10^{-4}$ for $a=1\,\mu\mathrm{m}$. Even using the larger effective interaction length $\lambda\approx 1\,\mathrm{nm}$ reported experimentally by~\citet{brown2014ionic}, one obtains $\lambda/a\approx 10^{-3}$, which remains small compared with the relevant gaps $\epsilon\approx 10^{-2}-10^{-1}$. Thus, the thin-layer approximation is appropriate for the neutral diffusiophoretic mechanism considered here.

Equations (\ref{laplace_dimensional}) -  (\ref{phoretic_slip}) form the basis of the diffusiophoresis mechanism, describing the solute concentration field, the fluid flow, and the coupling between them. 
In addition to the boundary conditions prescribed on the particle surface, we impose a no-slip condition for the fluid at the wall, together with an impermeability condition for the solute,

\begin{equation}\label{wall_bcs}
    \textbf{u} = \textbf{0} \quad \text{and} \quad \frac{\partial c}{\partial z} = 0 \quad \text{at} \quad z = 0.
\end{equation}

In what follows, we analyze a stationary particle and compute the hydrodynamic force and torque it experiences. For a freely suspended particle subject to zero net force and torque, the corresponding translational and rotational velocities are obtained via the standard resistance–mobility relations, which relate forces and torques to the translational and angular velocities of a spherical particle in the presence of a nearby wall. 

In this work, we consider a Janus particle with a large active cap and a small inert face, as shown in Fig.~\ref{fig:schematic}. In this case, the inert face is characterized by an angular extent $\phi$, 
which is assumed to be small, $\phi \ll 1$. 
As will become evident, our analysis is applicable to the complementary configuration in which the particle possesses a small active cap and a large inert face. { Indeed}, both surface distributions allow us to employ asymptotic methods and reveal that the small inert face or small active cap can have a pronounced effect on the motion of the Janus particle when brought sufficiently close to a wall.

We first consider the configuration in which the inert surface lies parallel to the wall, resulting in an axisymmetric flow. In Sec.~\ref{Tilted_section}, we extend the analysis to a slightly tilted inert surface, for which the flow becomes fully three  dimensional. Our focus is on the near-contact regime, where the gap distance $\epsilon a$ between the particle and the wall is small compared to the particle radius, $\epsilon \ll 1$. 

In this problem, there are two small parameters: the scaled separation distance $\epsilon$ and the inert-face angle $\phi$. When $\epsilon \ll 1$, the pressure is dominated by contributions from the narrow gap region. In this local geometry, the boundary of the sphere is parabolic, described by $h \approx h_0 \left(1 + r^2/(2ah_0)\right)$. Variations in the gap height then occur over the radial length scale $r \propto (ah_0)^{1/2}$, implying that the effective radial distance in the gap (normalized by $a$) is of order $\epsilon^{1/2}$. On the other hand, the inert face corresponds to the region with radial extent $r = a \sin{\phi} \approx a \phi$ at leading order. For this small-angle approximation to be quantitatively meaningful, we restrict attention to angles $\phi \lesssim 0.35$ rad, or equivalently $\phi \lesssim 20^\circ$. Over this range, the relative error in replacing $\sin\phi$ by $\phi$ remains approximately below $2\%$. 

Since there are two radial scalings, $\phi$ and $\epsilon^{1/2}$, we consider a distinguished limit when these scales are comparable, whereby both the active cap and the inert face contribute to the leading-order motion of the sphere within the lubrication approximation. This limit is written 
\begin{equation}\label{distinguished_limit_symmetric}
\phi \ll 1 \quad \text{and} \quad \epsilon \ll 1, \quad \text{with} \quad \Phi = \frac{\phi}{\epsilon^{1/2}} \quad \text{fixed}.
\end{equation}
For experimentally relevant Janus particle systems with colloid radii on the order of $1\text{--}5~\mu\mathrm{m}$, reported gap distances are typically about $200~\mathrm{nm}$~\citep{simmchen2016topographical, ketzetzi2020diffusion, ketzetzi2020slip}, corresponding to a representative dimensionless gap size of $\epsilon \approx 0.1$. In the distinguished limit considered here, $\phi \sim \epsilon^{1/2}$  gives
$
\phi \approx \sqrt{0.1} \approx 0.316~\mathrm{rad} \approx 18^\circ,
$
 which is consistent with the range $\phi \lesssim 20^\circ$ specified above.

\subsection{Dimensionless formulation}
We begin by outlining the dimensionless formulation of the problem at the particle scale before turning to the lubrication approximations. All lengths are scaled by the particle radius $a$. From the flux condition (\ref{activity}), the solute concentration scales as $a \alpha / D$. The phoretic slip condition (\ref{phoretic_slip}) then sets the characteristic velocity to be $\alpha \beta / D$. Finally, the Stokes equations (\ref{Stokes_equations}) yield the characteristic pressure scale $(\mu \alpha \beta)/ (a D)$ and {the  characteristic scale for the force $\textbf{f}$ of} $\mu a \alpha \beta/ D$. {We therefore define the dimensionless variables}, 

\begin{equation}\label{scaling_length_velocity_particle_scale}
R'  = \frac{r}{a},  \quad  Z'  =  \frac{z}{a},  \quad   C  =  \frac{Dc}{a \alpha}, \quad U_R'  = \frac{D u_r}{\alpha \beta }, \quad U_Z'  = \frac{D u_z}{\alpha \beta },  \quad P'  =  \frac{aDp}{\mu \alpha \beta}, \quad \textbf{F}'  =  \frac{D \textbf{f}}{\mu a \alpha \beta}.
\end{equation}
In these {variables}, the particle boundary is described by,
\begin{equation}\label{circle}
(Z' - 1-\epsilon)^2 + R'^2 = 1.
\end{equation}
Although we introduce the problem using particle-scale variables, it will become clear in the following sections that both the flow and solute transport are dominated by the narrow gap region. This localization is not a universal feature of self-diffusiophoresis, but is specific to models with prescribed flux boundary conditions (\ref{activity}). For kinetic models incorporating solute advection and reaction, however, the gap region alone does not determine the solute concentration, and a matched asymptotic analysis between the gap and particle-scale regions is necessary \cite{yariv2016wall,yariv2017boundary}. 

\subsection{Scalings in the lubrication limit}\label{lubrication_scaling}
We now rescale all relevant variables in the gap region, in line with the lubrication approximation. To motivate these scalings, we first examine how each variable depends on $\epsilon$. Within the gap region, the radial length scale as $R' = O(\epsilon^{1/2})$, while the gap height scales as $Z' = O(\epsilon)$. Moreover, since the tangent vector in the gap is approximately  $\mathbf{e}_r$, the phoretic slip condition (\ref{phoretic_slip}) suggests that the radial velocity scales as $U_R' = O\left(\epsilon^{-1/2}\right)$. The continuity equation (\ref{Stokes_equations}) then requires the vertical velocity to scale as $U_Z' = O(1)$. 

Finally, in the lubrication regime the flow is dominated by the balance between pressure gradients and viscous stresses, $\frac{\partial p}{\partial r} \sim \mu\frac{\partial^2 u_r}{\partial z^2}.$
This balance implies that the pressure scales as $P' = O\left(\epsilon^{-2}\right)$. {In the axisymmetric configuration, the force on the particle acts only in the $z$-direction, $\mathbf{F}' = F_Z' \mathbf{e}_z$, by symmetry. The pressure in the gap then leads to a vertical force scaling as $F_Z' = O(\epsilon^{-1})$.}

We define variables in the gap region using the stretched cylindrical coordinates,
\begin{equation}\label{scaling_length_velocity}
R  = \frac{R'}{\epsilon^{1/2}},  \quad  Z  =  \frac{Z'}{\epsilon}, \quad U_R  =  \epsilon^{1/2} U_R', \quad 
U_Z =U_Z', \quad P  =  \epsilon^{2} P', \quad F_Z  =   \epsilon F_Z'.
\end{equation}
In these coordinates, the particle boundary is described by a parabolic shape,
\begin{equation}\label{parabolic_profile}
Z = H(R) + O(\epsilon), \quad \text{where} \quad H(R) = 1 + \frac{R^2}{2}.
\end{equation}
The phoretic slip boundary condition on the particle surface (\ref{phoretic_slip}) becomes
\begin{equation}\label{phoretic_slip_dimensionless}
\textbf{U}(\textbf{x}) = \nabla_s C(\textbf{x}) \quad \text{at} \quad Z = H(R) + O(\epsilon),
\end{equation}
while the boundary condition for the solute flux on the particle surface (\ref{activity}) is
\begin{equation}\label{activity_dimensionless}
    -\textbf{n} \cdot \nabla C(\textbf{x})=
    \begin{cases}
        1,\quad \textbf{x} \in\text{ active cap,}\\
        0,\quad \textbf{x}\in\text{ inert face.}
    \end{cases}
\end{equation}

The interface between the active cap and the inert face is located at $r = a\sin{\phi} \approx a \phi$. Under the stretched coordinate introduced in (\ref{scaling_length_velocity}), at leading order in $\epsilon$, this boundary lies at $R = \Phi$. Thus, within the gap region, the active cap corresponds to $R > \Phi$, while the inert face corresponds to $R < \Phi$. Our approach is to solve the problem separately in these two regions, using a perturbation expansion in $\epsilon$, and then match the solutions by analyzing the transition layer near $R = \Phi$. 

\section{Axisymmetric Janus particle near a wall}\label{Axi_section}
\subsection{Solute concentration}\label{Solute_axi_section}
In the axisymmetric configuration, with the inert face oriented parallel to the wall, both the excess solute concentration and the flow are independent of $\theta$. In the gap region, (\ref{laplace_dimensional}) reduces to
\begin{equation}
\frac{\partial^2 C}{\partial Z^2} + \frac{\epsilon}{R}\frac{\partial}{\partial R}\left(R\frac{\partial C}{\partial R}\right) = 0.
\label{gap_laplace_axisymmetric}
\end{equation}

The solute flux conditions in the gap region~(\ref{activity_dimensionless}), written in the stretched coordinates, are
\begin{equation}
    -\textbf{n} \cdot \nabla C(R,Z)= \frac{1}{\epsilon}\left(\frac{\partial C}{\partial Z} - \epsilon R\frac{\partial C}{\partial R}\right)\left(1 + O(\epsilon)\right) = 
    \begin{cases}
        1,\quad \, R > \Phi + O(\epsilon)\\
        0,\quad \, R < \Phi + O(\epsilon)   
    \end{cases}
    \quad\text{at}\quad Z = H(R) + O(\epsilon).
    \label{gap_activity}
\end{equation}

To distinguish between the active and inert regions, we denote the excess solute concentration by $C_{+}(R,Z)$ for $R > \Phi$ and by $C_{-}(R,Z)$ for $R < \Phi$. Next, we write a regular perturbation expansion in $\epsilon$ for $C_{\pm}$,  
\begin{equation}
C_{\pm}(R,Z;\epsilon) = C_{\pm}^{(0)}(R,Z) + \epsilon C_{\pm}^{(1)}(R,Z) + \epsilon^2 C_{\pm}^{(2)}(R,Z) + ...,
\label{expansion_gap_scale}
\end{equation}
where the superscript indicates the corresponding order.
Our goal is to determine the leading order $C_{\pm}^{(0)}(R,Z)$. Substituting the expansion~(\ref{expansion_gap_scale}) into the gap-scale Laplace equation~(\ref{gap_laplace_axisymmetric}), with the impermeability condition at the wall~(\ref{wall_bcs}), shows that $C_{\pm}^{(0)} = C_{\pm}^{(0)}(R)$ is independent of $Z$. At leading order, the flux conditions~(\ref{gap_activity}) degenerate and therefore do not determine $C_{\pm}^{(0)}$. To resolve $C_{\pm}^{(0)}$, we proceed to the $O(\epsilon)$ problem and obtain solvability conditions. The Laplace equation at the $O(\epsilon)$ order is given by
\begin{equation}
\frac{\partial^2 C_{\pm}^{(1)}}{\partial Z^2} = -\frac{1}{R}\frac{d}{d R}\left(R \frac{dC_{\pm}^{(0)}}{dR}\right).
\label{first_order_laplace}
\end{equation}
Since $C_{\pm}^{(0)}$ is independent of $Z$, we integrate both sides of equation (\ref{first_order_laplace})  with respect to $Z$ and use the boundary condition on the wall (\ref{wall_bcs}) to find 
\begin{equation}
\frac{\partial C_{\pm}^{(1)}}{\partial Z} = -\frac{Z}{R}\frac{d}{d R}\left(R \frac{dC_{\pm}^{(0)}}{dR}\right).
\label{first_order_laplace_integrated}
\end{equation}

At first order, the flux condition~(\ref{gap_activity}) couples $C_{\pm}^{(1)}$ to $C_{\pm}^{(0)}$ through the relation
\begin{equation}
\frac{\partial C_{+}^{(1)}}{\partial Z} =  R\frac{d C_{+}^{(0)}}{d R}  + 1 \quad \text{and} \quad  \frac{\partial C_{-}^{(1)}}{\partial Z} = R\frac{d C_{-}^{(0)}}{d R}   \quad \text{at} \quad Z = H(R).
\label{gap_activity_first_order}
\end{equation}
Evaluating~(\ref{first_order_laplace_integrated}) at $Z = H$ and applying the flux condition~(\ref{gap_activity_first_order}) yields an ordinary differential equation for $C_{\pm}^{(0)}$,
\begin{equation}
\frac{H}{R}\frac{d}{d R}\left(R \frac{dC_{+}^{(0)}}{dR}\right) +  R\frac{d C_{+}^{(0)}}{d R}  + 1 = 0 \quad \text{and} \quad \frac{H}{R}\frac{d}{d R}\left(R \frac{dC_{-}^{(0)}}{dR}\right) +  R\frac{dC_{-}^{(0)}}{d R}  = 0.
\label{solvability_equation}
\end{equation}
These equations admit solutions for $C_{+}^{(0)}(R)$ and $C_{-}^{(0)}(R)$ such that
\begin{equation}
\frac{dC_{+}^{(0)}}{dR} = -\frac{1}{R} + \frac{K_{+}}{R H}
\quad \text{and} \quad
\frac{dC_{-}^{(0)}}{dR} = \frac{K_{-}}{R H},
\label{leading_gradient}
\end{equation}
where $K_{+}$ and $K_{-}$ are constants. Recall that $C_{-}^{(0)}$ refers to the domain $0 < R < \Phi$, whereas $C_{+}^{(0)}$ refers to $\Phi < R < \infty$. Accordingly, we require $C_{-}^{(0)}$ to remain bounded as $R \to 0$, while $C_{+}^{(0)}$ remains bounded as $R \to \infty$. From (\ref{leading_gradient}), we observe that near the origin $ C_{-}^{(0)} \sim K_{-}\log R.$
Boundedness as $R \to 0$ therefore requires $K_{-}=0$. 

The constant $K_{+}$, however, cannot be determined by the same argument, since the domain for $C_{+}^{(0)}$ does not contain the origin and $dC_0^{+}/dR$ decays to zero as $R \longrightarrow \infty$ regardless of the value of $K_{+}$. To obtain the value of $K_{+}$, an analysis on a transition region near $R = \Phi$ is needed to understand the transport of solute flux between the two regions. We claim that the analysis in the transition region requires the solute concentration gradient $\frac{dC^{(0)}}{dR}$ to be continuous at $R = \Phi$, see Appendix \ref{transition_region}. This condition yields
\begin{equation}
K_{+} = H(R = \Phi) = 1 + \frac{\Phi^2}{2}.
\label{constant_leading}
\end{equation}
In the next section, we use the resulting solute concentration gradient (\ref{leading_gradient}) to compute the flow field induced by the associated slip velocity.
\subsection{Flow analysis}\label{Flow_axi_section}
Similar to the expansion for the excess solute concentration in (\ref{expansion_gap_scale}), we expand the velocity and pressure in powers of $\epsilon$,
\begin{equation}
U_R(R,Z;\epsilon) = U_R^{(0)}(R,Z) + \epsilon U_R^{(1)}(R,Z) + ...   \quad \text{and} \quad  P(R,Z;\epsilon) = P^{(0)}(R,Z) + \epsilon P^{(1)}(R,Z) + ... .
\label{velocity_pressure_expansion}
\end{equation}
At leading order, the radial and axial components of the Stokes equations~(\ref{Stokes_equations}) reduce, respectively, to
\begin{equation}
\frac{\partial P^{(0)}}{\partial R} = \frac{\partial^2 U_R^{(0)}}{\partial Z^2}
\quad \text{and} \quad
\frac{\partial P^{(0)}}{\partial Z} = 0.
\label{Stokes_leading}
\end{equation}
Equation (\ref{Stokes_leading}) is commonly referred to as the lubrication approximation. The second equation implies that the leading-order pressure $P^{(0)}$ is independent of $Z$, and thus depends only on the radial coordinate, i.e., $P^{(0)}(R)$.

As noted above, at leading order the surface tangent in the gap region is approximately $\textbf{e}_r$; the phoretic slip boundary condition~(\ref{phoretic_slip_dimensionless}) for the radial velocity takes the form
\begin{equation}
U_R^{(0)} = \frac{dC^{(0)}}{dR} \quad \text{at} \quad Z = H(R).
\label{Stokes_bc}
\end{equation}
We integrate (\ref{Stokes_leading}) twice with respect to $Z$ and apply the boundary conditions at the particle surface~(\ref{Stokes_bc}) and the wall~(\ref{wall_bcs}) to obtain
\begin{equation}
U_R^{(0)} = \frac{1}{2}\frac{\partial P^{(0)}}{\partial R} Z(Z-H) +  \frac{Z}{H}\frac{dC^{(0)}}{dR}.
\label{velocity_axisymmetric}
\end{equation}

Rather than using the continuity equation directly, we instead adopt the mass conservation argument to proceed. In the axisymmetric gap region, since the particle is fixed, there can be no net flow across a radial cross-section; this implies that $\int_{0}^{H(R)} U_R^{(0)}(R,Z) \, dZ = 0.$ This constraint allows us to express the pressure gradient in terms of the solute concentration gradient,
\begin{equation}\label{leading_pressure_gradient}
 \frac{\partial P^{(0)}}{\partial R} = \frac{6}{H^2(R)}\frac{dC^{(0)}}{dR}.
\end{equation}
Since the $O(\epsilon^{-2})$ pressure in the gap must match the $O(1)$ pressure outside the gap, it follows that $P^{(0)} \longrightarrow 0$ as $R \longrightarrow \infty$. We can now integrate (\ref{leading_pressure_gradient}) once to write down the leading-order pressure explicitly, 
\begin{equation}
P^{(0)}(R) = -\int_{R}^{\infty} \frac{6}{H^2(\tilde{R})}\frac{dC^{(0)}(\tilde{R})}{d\tilde{R}} \: d\tilde{R}.
\label{pressure_formula_leading_integral}
\end{equation}

We note that in deriving (\ref{pressure_formula_leading_integral}), we have implicitly used the fact that $dC^{(0)}/dR$ is continuous across the entire domain $R \in (0,\infty)$, including at $R = \Phi$. The integral in (\ref{pressure_formula_leading_integral}) can then be evaluated exactly using the solute concentration gradient given in (\ref{leading_gradient}) and the constant determined in (\ref{constant_leading}),
\begin{equation}
P^{(0)}(R) = \begin{cases}
       \frac{6}{(R^2+2)^2} + 3\Phi^2\left(\frac{R^2+3}{(R^2+2)^2} + \ln{\left(\frac{R}{\sqrt{R^2+2}}\right)}\right),\quad \, R > \Phi,\\
       \frac{3(\Phi^2+1)}{\Phi^2+2} + 3\Phi^2 \ln{\left(\frac{\Phi}{\sqrt{\Phi^2+2}}\right)},\quad \, \qquad \quad \quad \quad R < \Phi.
    \end{cases}
    \label{pressure_formula}
\end{equation}

Finally, the leading-order vertical hydrodynamic force on the particle is dominated by the large pressure within the gap and, therefore, can be expressed using (\ref{pressure_formula}) as

\begin{equation}\label{leading_force}
F_Z^{(0)} = 2\pi \int_{0}^{\infty}RP^{(0)}(R) \: dR = \frac{6\pi}{2 + \Phi^2}. 
\end{equation}
In the limit $\Phi \ll 1$, the leading-order vertical hydrodynamic force approaches $3\pi$. Heuristically, this corresponds to the case where the inert portion of the particle is small compared to the lubrication region, so that the gap dynamics are dominated by the active face, see Fig. \ref{fig:vertical_velocity}. This result is consistent with the analysis of a fully active particle near a wall, as reported by~\citet{yariv2016wall,yariv2017boundary}.

\subsection{A complementary configuration}\label{Flipping_section}
It is useful to consider the complementary configuration, in which the Janus particle possesses a small active cap and a large inert face, with the active cap oriented toward the wall. In this geometry, the gap region $R < \Phi$ corresponds to the active cap, whereas $R > \Phi$ corresponds to the inert face. The analysis for this configuration follows the same framework as that developed in the previous section.

We denote by $\tilde{C}_{-}$ and $\tilde{C}_{+}$ the leading-order excess solute concentrations in the regions $R < \Phi$ and $R > \Phi$, respectively, and use a superscript tilde to distinguish all corresponding quantities in this complementary configuration. We obtain 
\begin{equation}\label{leading_gradient_flip}
    \frac{d\tilde{C}_{+}^{(0)}}{dR} =  \frac{\tilde{K}_{+}}{RH} \quad \text{and} \quad \frac{d\tilde{C}_{-}^{(0)}}{dR} =  -\frac{1}{R} + \frac{\tilde{K}_{-}}{RH},
\end{equation}
for some constants $\tilde{K}_{+}$ and $\tilde{K}_{-}$. Requiring the excess solute concentration to be bounded at the origin $R=0$ implies that $\tilde{K}_{-}=1$. An analysis of the transition region, analogous to that in Appendix \ref{transition_region} , shows that $d\tilde{C}^{(0)}/dR$ is continuous across $R=\Phi$, yielding
\begin{equation}\label{constant_leading_flip}
\tilde{K}_{+} = -\frac{\Phi^2}{2}.
\end{equation}
The pressure follows directly from the formula derived in (\ref{pressure_formula_leading_integral}),
\begin{equation}
\tilde{P}^{(0)}(R) = \begin{cases}
      -3\Phi^2\left(\frac{R^2+3}{(R^2+2)^2} + \ln{\left(\frac{R}{\sqrt{R^2+2}}\right)}\right),\quad \,\,\, \quad \quad  R > \Phi,\\
      \frac{6}{(R^2+2)^2}-\frac{3(\Phi^2+1)}{\Phi^2+2} -3\Phi^2\ln{\left(\frac{\Phi}{\sqrt{\Phi^2+2}}\right)},\quad \, R < \Phi.
    \end{cases}
    \label{pressure_formula_flip}
\end{equation}
The leading-order vertical hydrodynamic force (scaled by $\epsilon$) acting on the sphere due to this gap-scale region can now be obtained by integrating the pressure (\ref{pressure_formula_flip}),
\begin{equation}
\tilde{F}_Z^{(0)} = 2\pi \int_{0}^{\infty}R\tilde{P}^{(0)}(R) \: dR = \frac{3\Phi^2\pi}{2 + \Phi^2}.
\end{equation}

In the limit $\Phi \gg 1$, the leading-order vertical hydrodynamic force in this complementary configuration again approaches $3\pi$, mirroring the $\Phi \ll 1$ limit of the original configuration. This regime corresponds to the active portion of the particle being large compared to the lubrication region, so that the gap dynamics are dominated by the active face, despite the active cap itself being small. This result highlights that even a small active cap can generate a large force when the particle is in near contact with the wall, as the wall effectively interacts only with the active region of the Janus particle. This observation suggests that, within the fixed-flux model, near-wall phoretic repulsion or attraction could in principle be enhanced by optimizing the active-cap size relative to the lubrication gap scale. In practice, however, this conclusion may be modified by finite-rate surface kinetics, which can suppress solute accumulation in narrow gaps.

\begin{figure}
    \centering
    \includegraphics[width=0.9\columnwidth]{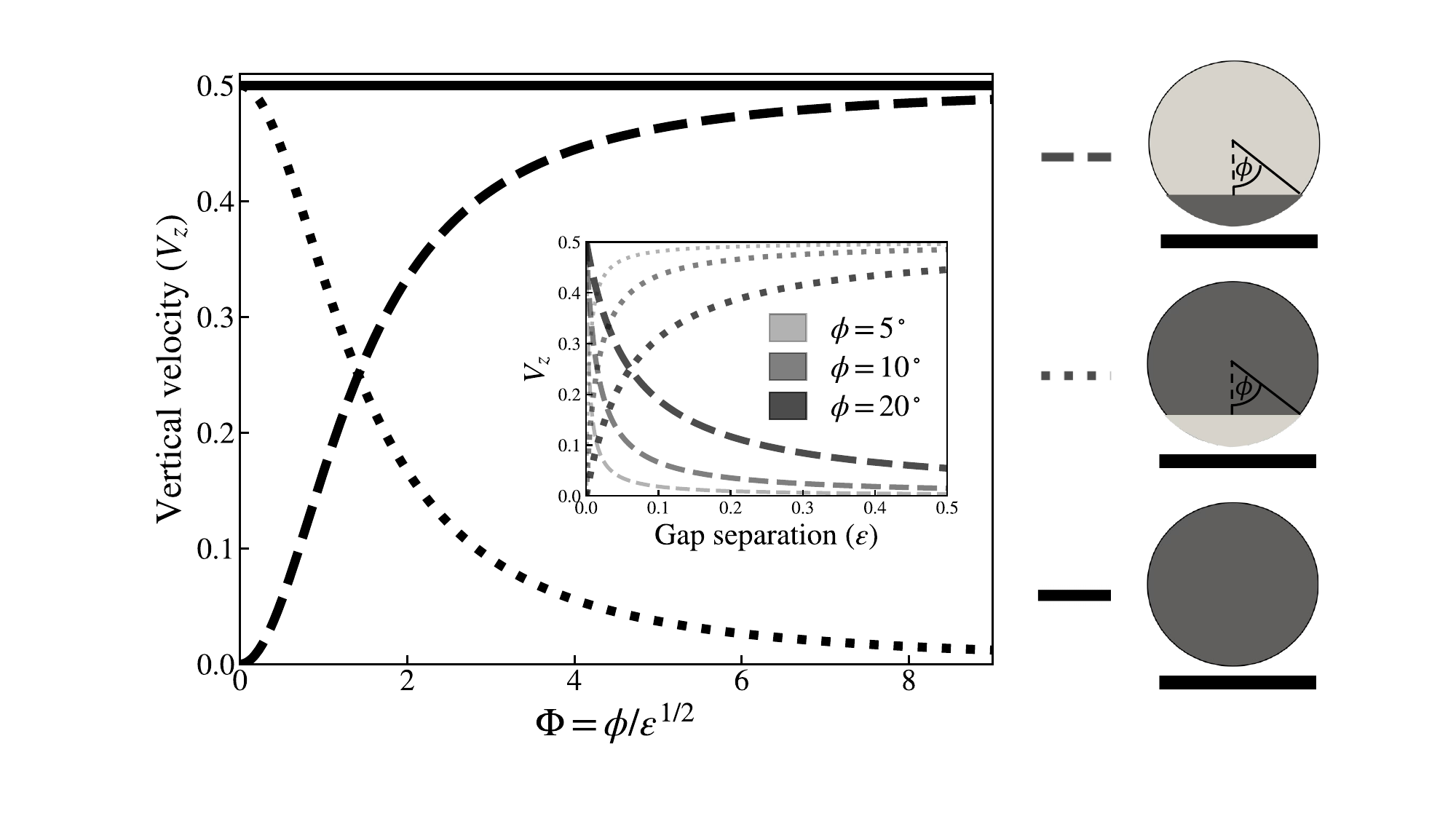}
    \caption{
       Translational velocity in the $z$-direction, normalized by $\alpha \beta /D$, of a freely-suspended Janus particle near a wall as a function of the parameter $\Phi = \phi/\epsilon^{1/2}$. The dotted curve corresponds to the original configuration, in which the particle has a large active cap and a small inert face, with the inert face oriented toward the wall. The dashed curve corresponds to the complementary configuration, where the particle has a small active cap and a large inert face, with the active cap oriented toward the wall. The velocity for a fully active particle (solid line) was obtained by~\citet{yariv2016wall,yariv2017boundary}. The inset shows the vertical velocity, normalized by $\alpha \beta /D$, as a function of the gap separation $\epsilon$, normalized by the particle radius $a$, for fixed cap angles $\phi=5^\circ$, $10^\circ$, and $20^\circ$; darker, thicker curves correspond to larger values of $\phi$.
    }
    \label{fig:vertical_velocity}
\end{figure}

So far, we have obtained the hydrodynamic force acting on a stationary Janus particle; these results can be used directly to compute the vertical rigid-body velocity $V_z$, normalized by $\alpha \beta / D$, of a freely-suspended Janus particle. This velocity is obtained by multiplying the dimensional force by the mobility coefficient $\epsilon/(6\pi\mu a)$ \cite{cox1967slow}. To leading order, we find that the vertical velocities for the original and complementary configurations approach the finite values
\begin{equation}
V_z \approx \frac{1}{2 + \Phi^2}
\quad \text{and} \quad
\tilde{V}_z \approx \frac{\Phi^2}{2(2 + \Phi^2)},
\label{phoretic_velocity_axisymmetric}
\end{equation}
respectively. We plot the vertical velocity $V_z$ and $\tilde{V}_z$ as a function of $\Phi$ for both the original configuration and the complementary configuration in Fig. \ref{fig:vertical_velocity}. Previous studies have reported hovering states for inert-facing Janus particles with large active caps~\citep{uspal2015self,ibrahim2016walls,mozaffari2016self}. The absence of a zero-crossing in Fig.~\ref{fig:vertical_velocity} does not contradict these earlier results, but instead indicates that hovering lies outside the asymptotic regime considered here, namely $\phi \ll 1$, $\epsilon \ll 1$, with $\Phi=\phi/\epsilon^{1/2}=O(1)$. In particular, hovering appears to occur either for order-one or larger separations, $\epsilon=O(1)$ or $\epsilon\gg1$, or, in the near-field regime $\epsilon\ll1$, for inert faces that are not asymptotically small. This interpretation is consistent with the computations of \citet{mozaffari2016self} (see their Fig.~12(f)), where near-field hovering with $\epsilon<1$ occurs for intermediate inert-face sizes, with angular extent approximately $24^\circ<\phi<35^\circ$; for smaller inert faces, hovering only occurs in the far field, $\epsilon>1$. Although a hovering state does not occur in the force-free setup considered here, Fig.~\ref{fig:vertical_velocity} suggests that hovering could arise through a balance with an external force, such as the particle's weight or the force due to a wall--particle repulsive potential.

\section{Slightly tilted Janus particle near a planar wall}\label{Tilted_section}
\begin{figure}
    \centering
    \includegraphics[width=0.70\columnwidth]{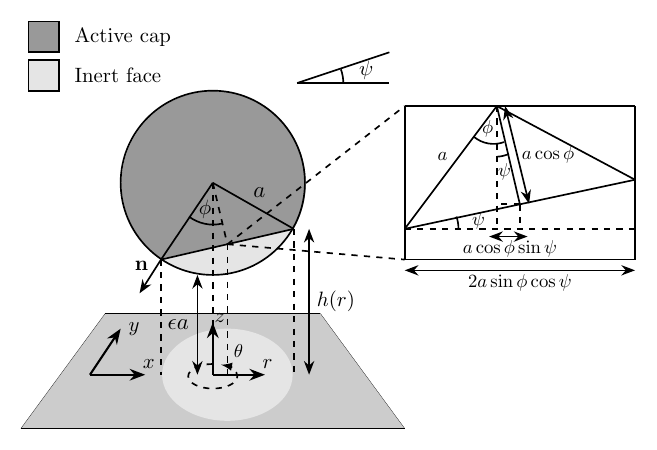}
    \caption{
       Schematic illustration of a spherical Janus particle of radius $a$ near a planar wall. The particle is tilted by a small angle $\psi$ about the $y$-axis relative to the wall.
    }
    \label{fig:schematic_tilted}
\end{figure}

In Sec.~\ref{Axi_section}, we analyzed a spherical Janus particle in the axisymmetric configuration where the inert surface of the particle is parallel to the wall. We now extend this analysis to the case in which the inert surface is slightly tilted about the $y$-axis, relative to the wall, by a small angle $\psi \ll 1$ (measured in radians); see Fig. \ref{fig:schematic_tilted}. It is convenient to adopt the same coordinate system as in the previous section, with the origin placed on the wall directly beneath the point on the particle surface closest to the wall. The gap height $h(r)$ is only a function of the radial coordinate. However, the inert face and the active cap now depend on the angular variable $\theta$. When projected onto the wall, the boundary between the active cap and the inert face forms an ellipse centered at $(r,\theta,z) = (a \cos\phi \sin\psi, 0, 0)$, with semi-axes of lengths $a \sin\phi \cos\psi$ in the $x$-direction and $a \sin\phi$ in the $y$-direction (see Fig.~\ref{fig:schematic_tilted}), which is described by the equation
\begin{equation}\label{ellipse}
    \left(\frac{r\sin{\theta}}{a \sin{\phi}}\right)^2 + \left(\frac{r\cos{\theta}-a\cos{\phi \sin{\psi}}}{a \sin{\phi} \cos{\psi}}\right)^2 = 1.
\end{equation}
Since both $\phi \ll 1$ and $\psi \ll 1$, equation~(\ref{ellipse}) may be simplified to
\begin{equation}\label{ellipse_approx}
\frac{r}{a} = \phi + \psi \cos\theta + O\left(\phi^2,\psi^2,\phi \psi\right).
\end{equation}
When the particle is not tilted ($\psi = 0$), this expression reduces to the result obtained in Sec.~\ref{Axi_section}.

As indicated in (\ref{distinguished_limit_symmetric}), we consider the distinguished limit comparing $\phi$ to $\epsilon^{1/2}$. Because the boundary between the active cap and the inert face \eqref{ellipse_approx} appears as a linear combination of $\phi$ and $\psi$, it is natural to adopt the same scaling for $\psi$; we therefore introduce
\begin{equation}
\Psi = \frac{\psi}{\epsilon^{1/2}} \in [0,\infty).
\end{equation}
In the gap-scale problem, to leading order in $\epsilon$, the boundary between the inert and active regions is given by $R = \Phi + \Psi \cos\theta$.
Although it is desirable to solve this problem for arbitrary $\Psi = O(1)$, we have found the general case analytically intractable. Nevertheless, progress can be made in the asymptotic limit $\Psi \ll 1$, or more precisely when $\Psi \ll \Phi$. In this limit, a domain perturbation approach may be employed, expanding the inert–active boundary about $R = \Phi$. Within this framework, the leading-order problem in $\Phi$ coincides with the solution obtained in Sec.~\ref{Axi_section}, and we focus on determining the $O(\Psi)$ correction. Since we restrict attention to $\phi \lesssim 20^\circ$, the condition $\psi \ll \phi$ suggests that the present approach is appropriate for tilt angles in the approximate range $0 \leq \psi \lesssim 10^\circ$, with improved accuracy as $\psi$ decreases.
\subsection{Solute concentration}\label{Solute_tilted_section}
In contrast to (\ref{gap_laplace_axisymmetric}), we retain the angular dependence in both the description of the flow and the solute concentration. In the gap region, (\ref{laplace_dimensional}) becomes
\begin{equation}\label{gap_laplace_tilted}
    \frac{\partial^2 C}{\partial Z^2} + \frac{\epsilon}{R}\frac{\partial}{\partial R}\left(R\frac{\partial C}{\partial R}\right) + \frac{\epsilon}{R^2}\frac{\partial^2 C}{\partial \theta^2} = 0.
\end{equation}

The solute flux conditions in the gap region~(\ref{activity_dimensionless}) take a similar form to (\ref{gap_activity}), since the normal vector to the particle surface has no angular dependence in our chosen coordinates, but with a modified boundary between the active cap and the inert face,
\begin{equation}
    -\textbf{n} \cdot \nabla C(R,\theta, Z)= \frac{1}{\epsilon}\left(\frac{\partial C}{\partial Z} - \epsilon R\frac{\partial C}{\partial R}\right)\left(1 + O(\epsilon)\right) = 
    \begin{cases}
        1,\quad \, R > \Phi + \Psi \cos{\theta} + O(\epsilon)\\
        0,\quad \, R < \Phi + \Psi \cos{\theta} + O(\epsilon)  
    \end{cases}
    \quad\text{at}\quad Z = H(R) + O(\epsilon).
    \label{gap_activity_tilted}
\end{equation}

As before, we denote the excess solute concentration by $C_{+}(R,\theta,Z)$ in the region $R > \Phi + \Psi \cos\theta$ and by $C_{-}(R,\theta,Z)$ in the region $R < \Phi + \Psi \cos\theta$. We then seek regular perturbation expansions of $C_{\pm}$ in $\epsilon$, 
\begin{equation}
C_{\pm}(R,\theta, Z;\epsilon, \Psi) = C_{\pm}^{(0)}(R,\theta, Z; \Psi) + \epsilon C_{\pm}^{(1)}(R,\theta, Z; \Psi) + \epsilon^2 C_{\pm}^{(2)}(R,\theta, Z; \Psi) + ...
\label{expansion_gap_scale_tilted}
\end{equation}
At this stage, no assumption has been made on the magnitude of $\Psi$ in constructing the expansion~(\ref{expansion_gap_scale_tilted}). We proceed by deriving the governing equations for $C_{\pm}^{(0)}$, following steps analogous to those leading to (\ref{solvability_equation}). We begin by noting from (\ref{gap_laplace_tilted}) that $C_{\pm}^{(0)} = C_{\pm}^{(0)}(R,\theta)$ is independent of $Z$. We then examine the $O(\epsilon)$ problem in (\ref{gap_laplace_tilted}), integrate once with respect to $Z$, and apply the flux condition~(\ref{gap_activity_tilted}) at $Z = H(R)$ to obtain
\begin{equation}
\frac{H}{R}\frac{\partial}{\partial  R}\left(R \frac{\partial C_{+}^{(0)}}{\partial R}\right) + \frac{H}{R^2}\frac{\partial^2 C_{+}^{(0)}}{\partial \theta^2} +R\frac{\partial C_{+}^{(0)}}{\partial R}  + 1 = 0 \quad \text{and} \quad \frac{H}{R}\frac{\partial}{\partial  R}\left(R \frac{\partial C_{-}^{(0)}}{\partial R}\right) + \frac{H}{R^2}\frac{\partial^2 C_{-}^{(0)}}{\partial \theta^2} +R\frac{\partial C_{-}^{(0)}}{\partial R}  = 0.
\label{solvability_equation_tilted}
\end{equation}
The equations in (\ref{solvability_equation_tilted}) alone do not determine the angular variations of $C_{\pm}^{(0)}$. A local analysis near the boundary $R = \Phi + \Psi \cos\theta$ is required to capture this dependency and relate $C_{+}^{(0)}$ to $C_{-}^{(0)}$. At leading order, neither the particle surface nor the wall contributes any concentration flux. Asymptotic matching with this local transition region near the boundary then necessitates that the \textit{normal} solute concentration flux across the boundary is conserved. Let $\tilde{\textbf{n}}(R,\theta)$ denote the normal vector to the curve $R = \Phi + \Psi \cos\theta$. We find
\begin{equation}
C_{+}^{(0)}(R,\theta; \Psi)  = C_{-}^{(0)}(R,\theta; \Psi) \quad \text{and}  \quad \tilde{\textbf{n}}\cdot \nabla C_{+}^{(0)}(R,\theta; \Psi) = \tilde{\textbf{n}}\cdot \nabla C_{-}^{(0)}(R,\theta; \Psi) \quad   \quad \text{at} \quad  R = \Phi+ \Psi \cos{\theta}.
\label{boundary_condition_tilted}
\end{equation}
A detailed justification of~\eqref{boundary_condition_tilted} is provided in Appendix~\ref{general_transition_region}.
It is difficult to solve equations (\ref{solvability_equation_tilted}) with constraints (\ref{boundary_condition_tilted}) and appropriate decays at the origin and in the far field. However, it is possible to obtain a perturbative solution in the limit when $\Psi \ll 1$. In particular, we seek regular expansions of $C_{\pm}^{(0)}$ in $\Psi$,
\begin{equation}
    C_{\pm}^{(0)}(R,\theta; \Psi) = C_{\pm}^{(00)}(R,\theta) + \Psi C_{\pm}^{(01)}(R,\theta) + \Psi^2 C_{\pm}^{(02)}(R,\theta) +...,
\label{second_expansion_gap_scale_tilted}
\end{equation}
where the second superscript indicates the order of the expansion in $\Psi$. We note that the subscripts plus ($+$) and minus ($-$) in (\ref{second_expansion_gap_scale_tilted}) correspond to solutions defined on the domains $\Phi < R < \infty$ and $0 \leq R < \Phi$, respectively, which differs from the notation adopted in (\ref{expansion_gap_scale_tilted}) to accommodate the domain perturbation in $\Psi$. The leading-order term $C_{\pm}^{(00)}$ in (\ref{second_expansion_gap_scale_tilted}) is equivalent to the axisymmetric solution obtained in (\ref{leading_gradient}). We now turn to determine $C_{\pm}^{(01)}$, which captures the leading-order effect of the tilt on the dynamics of the Janus particle.

To obtain the boundary conditions for $C_{\pm}^{(01)}$, we use a domain perturbation and first expand the normal solute concentration flux in (\ref{boundary_condition_tilted}) about $R=\Phi$ as
\begin{equation}\label{normal_flux_tilted}
    \tilde{\textbf{n}}\cdot \nabla C_{\pm}^{(0)}(R,\theta; \Psi) = \frac{\partial C_{\pm}^{(0)}}{\partial R} + \Psi\left(\cos \theta \frac{\partial^2 C_{\pm}^{(0)}}{\partial R^2} + \frac{\sin \theta}{R^2}  \frac{\partial C_{\pm}^{(0)}}{\partial \theta}  \right)  + O(\Psi^2) \quad \text{at} \quad R = \Phi.
\end{equation}
Substituting the expansion (\ref{second_expansion_gap_scale_tilted}) into (\ref{boundary_condition_tilted}) and (\ref{normal_flux_tilted}), and recalling that $C_{\pm}^{(00)} = C_{\pm}^{(00)}(R)$ is independent of $\theta$, we obtain
\begin{equation}
C_{+}^{(01)}(R,\theta)  = C_{-}^{(01)}(R,\theta) \quad \text{and} \quad \frac{\partial C_{+}^{(01)}(R,\theta)}{\partial R}  -  \frac{\partial C_{-}^{(01)}(R,\theta)}{\partial R} = \cos{\theta} \left(\frac{d^2 C_{-}^{(00)}(R)}{d R^2}  -  \frac{d^2 C_{+}^{(00)}(R)}{d R^2} \right)\quad \text{at} \quad R = \Phi.
\label{first_boundary_condition_tilted}
\end{equation}
Motivated by the boundary conditions (\ref{first_boundary_condition_tilted}), it is natural to seek the angular dependence of $C_{\pm}^{(01)}$ by assuming the form
\begin{equation}
C_{\pm}^{(01)}(R,\theta)  = \mathcal{C}_{\pm}(R) \cos{\theta}.
\label{ansatz}
\end{equation}
Under this ansatz, (\ref{first_boundary_condition_tilted}) is transformed into boundary conditions depending solely on the radial coordinate,
\begin{equation}
\mathcal{C}_{+}(R)  = \mathcal{C}_{-}(R) \quad \text{and} \quad  \frac{d \mathcal{C}_{+}(R)}{dR}  -  \frac{d\mathcal{C}_{-}(R)}{d R} =  \frac{d^2 C_{-}^{(00)}(R)}{d R^2}  -  \frac{d^2 C_{+}^{(00)}(R)}{d R^2} = \frac{1}{H(R = \Phi)} \quad \text{at} \quad R = \Phi,
\label{reduced_first_boundary_condition_tilted}
\end{equation}
where we have substituted the axisymmetric solution (\ref{leading_gradient}) for simplifying (\ref{reduced_first_boundary_condition_tilted}) and $H(R = \Phi)=1+\frac{\Phi^2}{2}$. On the other hand, from (\ref{solvability_equation_tilted}), the governing equations for $\mathcal{C}_{\pm}$ are simplified to
\begin{equation}
\frac{H}{R}\frac{d}{d R}\left(R \frac{d\mathcal{C}_{\pm}}{dR}\right)   - \frac{H\mathcal{C}_{\pm}}{R^2} + R\frac{d\mathcal{C}_{\pm}}{d R}= 0.
\label{solvability_tilted}
\end{equation}
Note that both $\mathcal{C}_{+}$ and $\mathcal{C}_{-}$ satisfy the same homogeneous differential equation, since the inhomogeneous term in (\ref{solvability_equation_tilted}) appears only at leading order for $C_{\pm}^{(00)}$. Nevertheless, the boundary conditions (\ref{reduced_first_boundary_condition_tilted}) provide the necessary inhomogeneity to specify the solutions $\mathcal{C}_{\pm}$. We solve $\mathcal{C}_{-}$ in the domain $R \in [0,\Phi)$ and $\mathcal{C}_{+}$ in the domain $R \in (\Phi,\infty)$. Accordingly, $\mathcal{C}^{-}$ is the solution of (\ref{solvability_tilted}) that remains regular at the origin, while $\mathcal{C}^{+}$ is the solution that is regular in the far field. We find that the general forms of the solutions satisfying these decay conditions are
\begin{equation}
\mathcal{C}_{-}(R) = \mathcal{K}_{-} R\,{}_2F_1\left(1-\frac{1}{\sqrt{2}}, 1 + \frac{1}{\sqrt{2}}; 2 ; -\frac{R^2}{2}\right) \quad \text{and} \quad \mathcal{C}_{+}(R) = \mathcal{K}_{+} R^{-\left(1+\sqrt{2}\right)}{}_2F_1\left(\frac{1}{\sqrt{2}}, 1+\frac{1}{\sqrt{2}}; 1+\sqrt{2} ; -\frac{2}{R^2}\right),
\label{solute_concentration_tilted}
\end{equation}
where ${_2F_1}$ is the hypergeometric function. The constants $\mathcal{K}_{-}$ and $\mathcal{K}_{+}$ in (\ref{solute_concentration_tilted}) are determined by the boundary conditions (\ref{reduced_first_boundary_condition_tilted}). 

\subsection{Flow analysis}\label{Flow_tilted_section}
In the previous section, we derived the excess solute concentration using a double perturbation expansion in $\epsilon$ and $\Psi$. We now extend this framework to determine the corresponding velocity and pressure fields, focusing on the leading order in $\epsilon$ and the first-order correction in $\Psi$, i.e., the $(01)$-order in the notation of the preceding section. We write
\begin{subequations}
\begin{align}
    U_{\{R,\theta \}}(R,\theta,Z; \epsilon, \Psi) &= U_{\{R,\theta \}}^{(00)}(R,\theta,Z) +  \Psi U_{\{R,\theta \}}^{(01)}(R,\theta,Z) + O\left(\epsilon, \Psi^2\right), \\
P(R,\theta,Z; \epsilon, \Psi) &= P^{(00)}(R,\theta,Z) +  \Psi P^{(01)}(R,\theta,Z) + O\left(\epsilon, \Psi^2\right), 
\end{align}
\end{subequations}
where $U_{R}^{(00)}$ and $P^{(00)}$ correspond to the solutions of the axisymmetric problem given by (\ref{velocity_axisymmetric}) and (\ref{pressure_formula}),  respectively. 

Since the solute concentration and the velocity field are coupled through the phoretic slip boundary condition (\ref{phoretic_slip_dimensionless}), we choose the angular dependence of the velocity and pressure fields based on that of the solute concentration in (\ref{ansatz}). In particular, we let
\begin{equation}
U_R^{(01)}(R,\theta,Z) = \mathcal{U}_R(R,Z)\cos \theta, \quad  U_\theta^{(01)}(R,\theta,Z) = \mathcal{U}_\theta(R,Z)\sin \theta, \quad \text{and} \quad P^{(01)}(R,\theta,Z) = \mathcal{P}(R,Z) \cos \theta.
\label{velocity_pressure_ansatz}
\end{equation}
These assumptions allow us to simplify the lubrication approximation of the Stokes equations (\ref{Stokes_equations}) to a system of differential equations involving only the radial and axial coordinates,
\begin{equation}
\frac{\partial \mathcal{P}}{\partial R} = \frac{\partial^2 \mathcal{U}_R}{\partial Z^2}, \quad -\frac{\mathcal{P}}{R} = \frac{\partial^2 \mathcal{U}_\theta}{\partial Z^2}, \quad \text{and} \quad \frac{\partial \mathcal{P}}{\partial Z} = 0, 
\label{Stokes_tilted}
\end{equation}
together with slip boundary conditions imposed at the particle surface,
\begin{equation}
\mathcal{U}_R = \frac{d\mathcal{C}}{dR} \quad \text{and} \quad \mathcal{U}_\theta = -\frac{\mathcal{C}}{R} \quad \text{at} \quad Z = H(R).
\label{phoretic_slip_tilted}
\end{equation}
Notice from (\ref{Stokes_tilted}) that the pressure $\mathcal{P}$ is independent of $Z$. We integrate (\ref{Stokes_tilted}) twice with respect to $Z$ and apply the boundary conditions at the particle surface (\ref{phoretic_slip_tilted}) and at the wall (\ref{wall_bcs}) to obtain 
\begin{subequations}
\begin{align}
\mathcal{U}_R(R,Z) &= \frac{1}{2}\frac{d\mathcal{P}}{dR}(Z^2-ZH) + \frac{Z}{H} \frac{d\mathcal{C}}{dR}, \label{v_r_tilted} \\
\mathcal{U}_\theta(R,Z) &= -\frac{1}{2}\frac{\mathcal{P}}{R}(Z^2-ZH) - \frac{Z}{H} \frac{\mathcal{C}}{R}. \label{v_theta_tilted}
\end{align}
\end{subequations}
By integrating the continuity equation (\ref{Stokes_equations}) across the gap, we obtain an integral form of the mass conservation equation,
\begin{equation}
\frac{d}{dR} \left(R \int_{0}^{H(R)} \mathcal{U}_R(R,Z) \: dZ\right) + \int_{0}^{H(R)} \mathcal{U}_\theta(R,Z) \: dZ = 0.
\label{continuity_integral}
\end{equation}
Substituting (\ref{v_r_tilted}) and (\ref{v_theta_tilted}) into (\ref{continuity_integral}) yields a differential equation for the pressure $\mathcal{P}$,
\begin{align}
\frac{d}{dR}\left(RH^3 \frac{d\mathcal{P}}{dR}\right)-\frac{H^3 \mathcal{P}}{R} &= 6\left(\frac{d}{dR}\left(RH \frac{d\mathcal{C}}{dR}\right) - \frac{H \mathcal{C}}{R}\right) \label{Reynolds_eq_exact_derivative}  
\\
&
= 6R\left(\frac{H}{R}\frac{d}{d R}\left(R \frac{d\mathcal{C}}{dR}\right) - \frac{H\mathcal{C}}{R^2} + R\frac{d\mathcal{C}}{d R}\right). \label{Reynolds_eq_expand}
\end{align}
Thus far, we have derived the Reynolds equation for the pressure (\ref{Reynolds_eq_expand}), without using any specific properties of the excess solute concentration $\mathcal{C}$. We now note that the right-hand side of (\ref{Reynolds_eq_expand}) is the same expression as the left-hand side of (\ref{solvability_tilted}), up to a multiplicative factor of $6R$. Consequently, it vanishes in the regions $R \in [0,{\Phi})$ and $R \in (\Phi,\infty)$, away from the discontinuity in $d\mathcal{C}/dR$ at $R=\Phi$. We therefore introduce $\mathcal{P}_{-}$ and $\mathcal{P}_{+}$ to denote the pressure in the domains $R \in [0,\Phi)$ and $R \in (\Phi,\infty)$, respectively. With this notation, equation (\ref{Reynolds_eq_expand}) may be written as
\begin{equation}
    \frac{d}{dR}\left(RH^3 \frac{d\mathcal{P}_{\pm}}{dR}\right)-\frac{H^3 \mathcal{P}_{\pm}}{R} = 0.
    \label{pressure_tilted}
\end{equation}
Since $\mathcal{P}_{-}$ and $\mathcal{P}_{+}$ are the solutions of (\ref{pressure_tilted}) satisfying conditions, respectively, of regularity at the origin and decay in the far field,  we obtain
\begin{equation}
\mathcal{P}_{-}(R) = \mathcal{Q}_{-} R{}_2F_1\left(2-\frac{\sqrt{10}}{2}, 2 + \frac{\sqrt{10}}{2}; 2 ; -\frac{R^2}{2}\right) \quad \text{and} \quad \mathcal{P}_{+}(R) = \mathcal{Q}_{+} R^{-(3+\sqrt{10})}{}_2F_1\left(1+\frac{\sqrt{10}}{2}, 2+\frac{\sqrt{10}}{2}; 1+\sqrt{10} ; -\frac{2}{R^2}\right),
\label{pressure_formula_tilted}
\end{equation}
where $\mathcal{Q}_{\pm}$ are unknown constants. Similar to the regularity conditions for $\mathcal{C}$ at $R = \Phi$ in (\ref{reduced_first_boundary_condition_tilted}), we require the pressure $\mathcal{P}$ to be continuous at $R=\Phi$, 
\begin{equation}
\mathcal{P}_{+}(R) = \mathcal{P}_{-}(R) \quad \text{at} \quad R = \Phi,
\label{pressure_tilted_continuous}
\end{equation}
whereas the corresponding gradients, $d\mathcal{P}_{\pm}/dR$, inherit a jump discontinuity due to the discontinuity in the solute concentration gradient $d\mathcal{C}/dR$ at $R=\Phi$. To determine the magnitude of the jump in the pressure gradient, we interpret (\ref{Reynolds_eq_exact_derivative}) in the weak sense and integrate across $R=\Phi$, from $R=\Phi^{-}$ to $R=\Phi^{+}$, yielding
\begin{equation}
\frac{d \mathcal{P}_{+}(R)}{dR}  -  \frac{d\mathcal{P}_{-}(R)}{d R}  = \frac{6}{H^2}\left(\frac{d \mathcal{C}_{+}(R)}{dR}  -  \frac{d\mathcal{C}_{-}(R)}{d R}\right)  = \frac{6}{H^3(R = \Phi)}\quad \text{at} \quad R = \Phi,
\label{pressure_tilted_jump}
\end{equation}
where the last equation is obtained by substituting in (\ref{reduced_first_boundary_condition_tilted}). In deriving (\ref{pressure_tilted_jump}), we have implicitly used the continuity of both $\mathcal{P}$ and $\mathcal{C}$ at $R=\Phi$, so that the integrals of $H^3 \mathcal{P}/R$ and $H\mathcal{C}/R$ across $R=\Phi$ vanish. The boundary conditions (\ref{pressure_tilted_continuous}) and (\ref{pressure_tilted_jump}) specify the value of the constants $\mathcal{Q}_{-}$ and $\mathcal{Q}_{+}$.
\subsection{Phoretic motion}\label{Tilted_motion_section}
Thus far, we have obtained the leading-order effects of the tilt on the velocity field \eqref{v_r_tilted}–\eqref{v_theta_tilted} and on the pressure \eqref{pressure_formula_tilted}. We now use these results to compute the hydrodynamic force in the $x$-direction, $F_x$, normalized by $(\mu a \alpha \beta)/(D\epsilon)$, and the torque in the $y$-direction, $T_y$, normalized by $(\mu a^2 \alpha \beta)/(D\epsilon)$. Note that in the axisymmetric case, both quantities vanish by symmetry. The force in the $z$-direction is not computed here, since its leading-order contribution coincides with that of the axisymmetric case \eqref{leading_force}, with the tilt introducing only an $O(\Psi)$ correction. We expand $F_x$ and $T_y$ as
\begin{equation}
    F_x = \Psi F_x^{(01)} + O\left(\epsilon, \Psi^2\right) \quad \text{and} \quad T_y = \Psi T_y^{(01)} + O\left(\epsilon, \Psi^2\right).
\label{force_torque_expansion}
\end{equation}
Since the $(01)$-order velocity field and pressure have the same form as those for a sphere translating parallel near a plane wall, $F_x^{(01)}$ and $T_y^{(01)}$ can be readily computed using a known lubrication-theory analysis \citep{goldman1967slow},
\begin{subequations}
\begin{align}
F^{(01)}_x  &= -\pi \int_{0}^{\infty} R\left(\frac{\partial \mathcal{U}_R}{\partial Z} - \frac{\partial \mathcal{U}_\theta}{\partial Z} + R\mathcal{P} \right)_{Z = H(R)} \: dR, \label{force_tilted}\\
T^{(01)}_y  &= \pi \int_{0}^{\infty} R\left(\frac{\partial \mathcal{U}_R}{\partial Z} - \frac{\partial \mathcal{U}_\theta}{\partial Z} \right)_{Z = H(R)}  \: dR. \label{torque_tilted}
\end{align}
\end{subequations}
In the near-contact regime, the translational motion parallel to the wall and the rotational dynamics are coupled, so that the directions of the force and torque do not translate directly to the direction of particle motion. Rather than presenting $F^{(01)}_x$ and $T^{(01)}_y$ at this stage, we continue our calculation for the rigid-body velocities of a freely-suspended Janus particle, obtained from the force- and torque-free conditions. We denote $V_x$, normalized by $(\alpha \beta)/(D\epsilon)$, and $\Omega_y$, normalized by $(\alpha \beta)/(aD\epsilon)$, as the particle’s translational velocity in the $x$-direction and angular velocity about the $y$-axis, respectively. The force and torque are related to the velocities by the following relation \citep{goldman1967slow},
\begin{equation}
\begin{pmatrix}
F_x \\
T_y
\end{pmatrix}
\approx
\frac{4 \pi}{5}\log{(1/\epsilon)}
\begin{pmatrix}
4 & -1 \\
-1 & 4
\end{pmatrix}
\begin{pmatrix}
V_x \\
\Omega_y
\end{pmatrix},
\label{resistance_matrix}
\end{equation}
where only the leading-order contributions in the near-contact limit are retained. Inverting the resistance relation (\ref{resistance_matrix}) allows the velocities to be expressed in terms of the force and torque; keeping only the leading-order contribution due to the tilt, as in (\ref{force_torque_expansion}), we obtain
\begin{equation}
\begin{pmatrix}
V_x \\
\Omega_y
\end{pmatrix}
\approx
\frac{\Psi}{12 \pi \log{(1/\epsilon)}}
\begin{pmatrix}
4 & 1 \\
1 & 4
\end{pmatrix}
\begin{pmatrix}
F_x^{(01)} \\
T_y^{(01)}
\end{pmatrix}.
\label{mobility_matrix}
\end{equation}

\begin{figure}
    \centering
    \includegraphics[width=0.8\columnwidth]{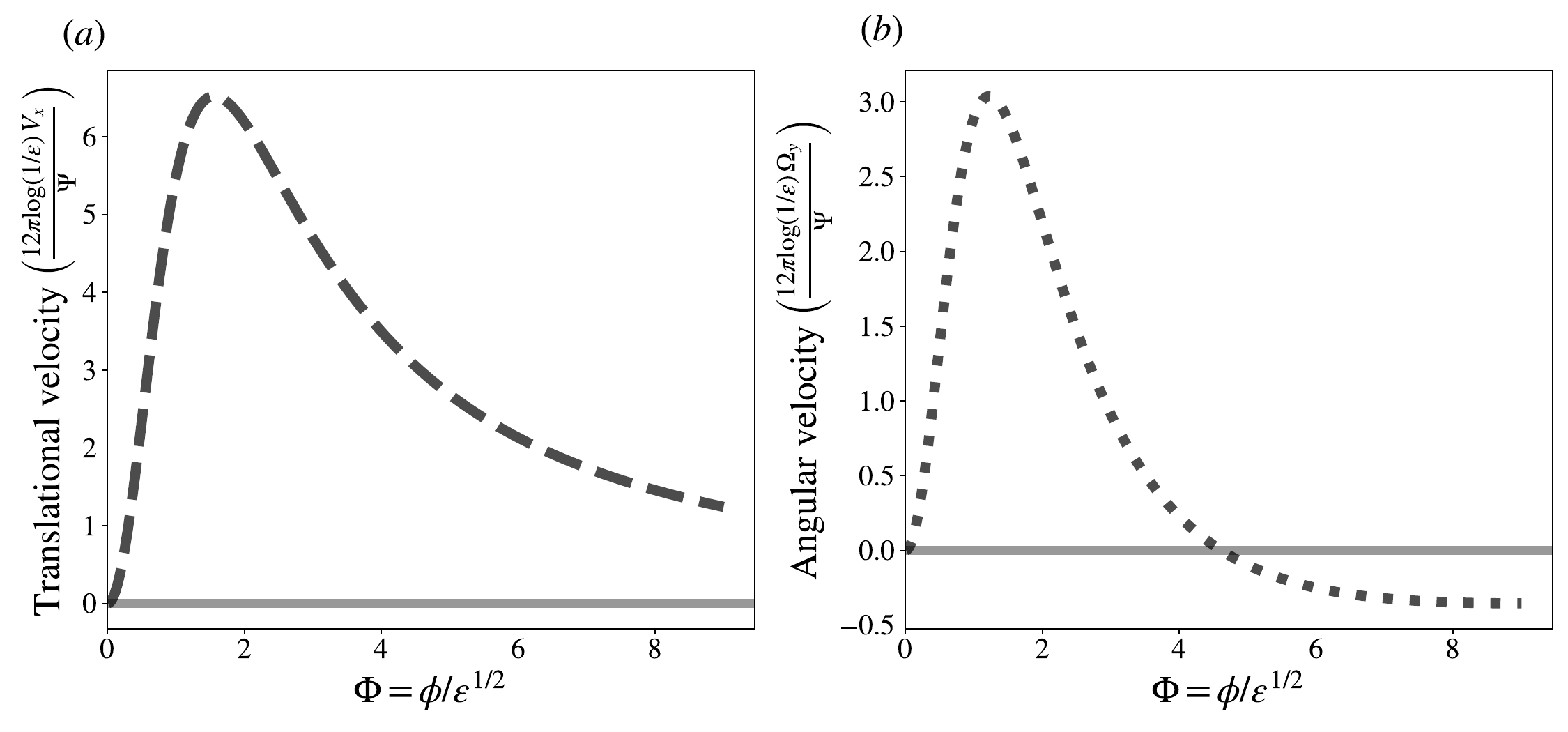}
    \caption{
       (a) Translational velocity in the $x$-direction of a slightly tilted, freely-suspended Janus particle near a wall, plotted as a function of the parameter $\Phi = \phi/\epsilon^{1/2}$. (b) Angular velocity around the $y$-axis of a slightly tilted, freely-suspended Janus particle near a wall, plotted as a function of the parameter $\Phi = \phi/\epsilon^{1/2}$.
    }
    \label{fig:force_torque}
\end{figure}

In what follows, we present numerical results for $V_x$ and $\Omega_y$ for different values of $\Phi$. For each $\Phi$, we first solve a linear system for the constants $\mathcal{K}_{-}$ and $\mathcal{K}_{+}$ in the excess solute concentration $\mathcal{C}_{\pm}$, as defined in (\ref{solute_concentration_tilted}), using the boundary conditions (\ref{reduced_first_boundary_condition_tilted}). We then determine the constants $\mathcal{Q}_{-}$ and $\mathcal{Q}_{+}$ for the pressure $\mathcal{P}_{\pm}$, as defined in (\ref{pressure_formula_tilted}), by solving a second linear system based on the boundary conditions (\ref{pressure_tilted_continuous}) and (\ref{pressure_tilted_jump}). With the excess solute concentration $\mathcal{C}$ and pressure $\mathcal{P}$ in hand, we compute the velocities using (\ref{v_r_tilted}) and (\ref{v_theta_tilted}) and numerically integrate the force and torque from (\ref{force_tilted}) and (\ref{torque_tilted}). Finally, multiplying the resulting force and torque by the mobility coefficients in (\ref{mobility_matrix}) yields the particle’s phoretic velocities as shown in Fig. \ref{fig:force_torque}.

\begin{figure}
    \centering
    \includegraphics[width=0.7\columnwidth]{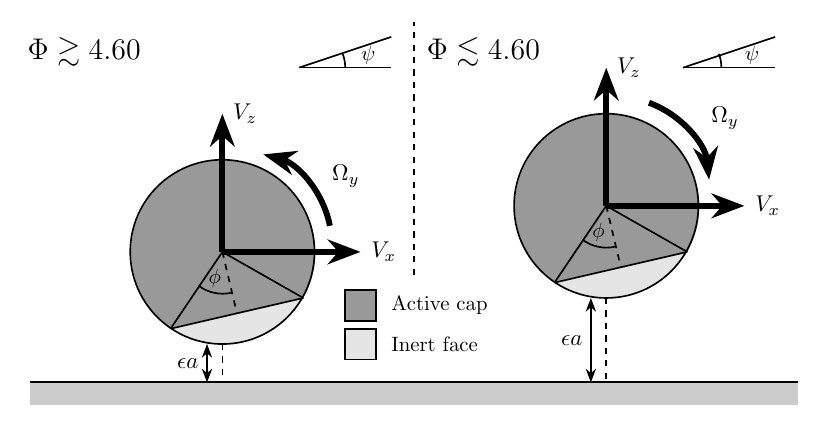}
    \caption{
       Schematic illustration of a slightly tilted Janus particle of radius $a$ near a planar wall, indicating the direction of its motion. For all values of $\Phi = \phi / \epsilon^{1/2}$, the particle translates in the positive $x$- and $z$-directions. In contrast, the direction of rotation depends on $\Phi$: \emph{(left)} for $\Phi \gtrsim 4.6$, the particle rotates counterclockwise, whereas \emph{(right)} for $\Phi \lesssim 4.6$, it rotates clockwise.
    }
    \label{fig:phoretic_motion}
\end{figure}

Although the translational velocity in the $x$-direction remains positive for all values of $\Phi$, the angular velocity changes sign at $\Phi \approx 4.60$. This behavior is striking, as it implies a change in the rotational stability of the particle. For a Janus particle with fixed cap size $\phi$, when the particle is sufficiently far from the wall such that $\phi / \epsilon^{1/2} < 4.60$, a small tilt in its orientation produces a restoring rotation that drives the particle back toward the axisymmetric configuration. In contrast, when the particle is extremely close to the wall so that $\phi / \epsilon^{1/2} > 4.60$, a small tilt induces a destabilizing rotation, causing the particle to rotate away from the axisymmetric state. This phoretic motion of the Janus particle is illustrated in Fig. \ref{fig:phoretic_motion}. 

It is important to note that the directions of motion predicted here are in qualitative agreement with the bi-spherical coordinate analysis of ~\citet{mozaffari2016self}, although a detailed comparison is limited because our study focuses on the slightly tilted regime. \citet{mozaffari2016self} observed that swimmers with larger active areas tend to rotate clockwise as they approach a wall, whereas at very small separations the torque reverses and counterclockwise rotation dominates (see discussions on page 20 or Fig. 10 of \citet{mozaffari2016self}). These remarks are consistent with our analysis. However, \citet{mozaffari2016self} did not comment on how these changes in the direction of motion arise in the lubrication regime, nor did they discuss how the directions depend on the parameter $\Phi = \phi/\epsilon^{1/2}$, which relates the cap size to the gap distance.

Similar to the analysis in Sec.~\ref{Flipping_section}, our framework in this section also applies to the complementary configuration consisting of a small active cap and a large inert face, with the active cap oriented toward the wall. We find that the direction of motion in this complementary configuration is nearly opposite to that of the original one. Specifically, when the particle is close (but not extremely close) to the wall, it rotates counterclockwise, whereas in the extreme near-wall limit it rotates clockwise. The translational motion parallel to the wall, $V_x$, is in the negative-$x$ direction, while the vertical velocity $V_z$ has the same direction as in the original configuration, albeit with a different magnitude, as noted in Sec.~\ref{Flipping_section}.

\section{Conclusion}\label{conclusion}
In this study, we examined the dynamics of a Janus particle near a solid wall. For an axisymmetric configuration in which the inert face is oriented parallel to the boundary, we obtained the vertical velocity of a freely suspended Janus particle for both the original and complementary configurations, as illustrated in Fig.~\ref{fig:vertical_velocity}. In the near-contact regime, the hydrodynamic force is dominated by the lubrication force within the narrow particle–wall gap and is determined by the extent to which the wall effectively “sees” the active portion of the particle; even a small inert face or, conversely, a small active cap can have a pronounced influence on the particle’s motion when it is brought sufficiently close to the wall. This behavior is captured by the dimensionless parameter $\Phi = \phi / \epsilon^{1/2}$, which compares the cap size to the square root of the gap distance.

Although the analysis of the axisymmetric configuration in Sec.~\ref{Axi_section} is relatively straightforward, the methods developed to treat mixed boundary conditions in the lubrication regime, detailed in Appendix~\ref{transition_region} and Appendix~\ref{general_transition_region}, are broadly applicable and readily extensible to more complex boundary conditions, such as those arising from periodically patterned Janus particles or, more generally, lubrication flows with mixed boundary conditions. We are not aware of prior work that establishes the continuity of the solute concentration and its first derivative through matched-asymptotic arguments; the present framework therefore provides a systematic asymptotic basis for identifying regularity of key physical quantities in related problems. Several natural extensions of the present framework also remain open. In particular, spatially varying surface mobility or activity, as well as osmotically driven flows along the bounding substrate, could be incorporated in the axisymmetric setting through modifications to the slip or solute-flux boundary conditions. These extensions should remain analytically tractable and provide useful directions for future work.

When the particle is slightly tilted, with tilt angle $\psi \ll 1$, the solute concentration and flow fields become fully three-dimensional. We therefore employ a double perturbation expansion in $\epsilon$ and $\Psi = \psi / \epsilon^{1/2}$ to determine the leading-order effects of the tilt on the particle’s motion. While the translational velocity parallel to the wall remains positive for all values of $\Phi$, the rotational velocity about the tilt axis changes sign at a finite value of $\Phi$. This direction of motion is qualitatively consistent with the findings of \citet{mozaffari2016self} and demonstrates that the axisymmetric configuration may be either rotationally stable or unstable, depending on the particle–wall separation in the near-contact regime. Although \citet{mozaffari2016self} did not analyze this stability in detail, it would be valuable to adapt their bispherical-coordinate analysis, or alternatively to use finite-element methods, to examine more closely the distinguished limit considered in the present study and validate the asymptotic results a posteriori.

One limitation of the present work is the restriction to small tilt angles, $\Psi \ll 1$. This assumption, however, should be interpreted as enabling a local stability analysis about a nearly axisymmetric state, rather than as a general description of Janus particles at arbitrary orientations. Previous studies have reported that Janus particles may skate along a wall at a finite, steady tilt angle~\citep{uspal2015self, ibrahim2016walls, mozaffari2016self, turk_singh_stone_2025}. Our analysis cannot capture this behavior, as the tilt introduces only an $O(\Psi)$ correction to the $O(1)$ velocity in the normal direction to the wall given by~(\ref{phoretic_velocity_axisymmetric}). A full solution of~(\ref{solvability_equation_tilted}) subject to the boundary conditions~(\ref{boundary_condition_tilted}) in the regime $\Psi = O(1)$ would determine whether steady skating states occur in the lubrication limit. Although the calculations are not presented here, we have analyzed the two-dimensional counterpart of the Janus particle–wall problem and found that steady skating does not occur for any finite tilt angle $\Psi = O(1)$ in the lubrication limit. This suggests that such skating behavior may be intrinsic to the three-dimensional setting. 

A second caveat is that Brownian motion is neglected. Brownian motion may be important for experimentally relevant Janus particles, especially in the bulk for particle radii of order $1$--$5\,\mu\mathrm{m}$. In the lubrication regime considered here, however, its effect on motion in the direction normal to the wall is expected to be reduced. The Brownian noise amplitude scales with the square root of the diffusion coefficient, which itself scales inversely with the hydrodynamic resistance. Since the lubrication resistance diverges near the wall, wall-normal Brownian fluctuations are suppressed at small gaps. Rotational diffusion may nevertheless remain important, as it can perturb a tilted particle away from axisymmetric orientations and reduce near-wall retention times, as reported by \citet{mozaffari2016self}.

The present lubrication analysis also highlights a potential shortcoming of the zeroth-order kinetic model. 
Although the particle is force-free, the individual slip-driven and motion-induced lubrication forces each scale as $O(\epsilon^{-1})$, while the particle velocity remains $O(1)$ (see Sec.~\ref{lubrication_scaling}). 
The associated mechanical power therefore scales as $O(\epsilon^{-1})$ and diverges as $\epsilon \to 0$. 
Since a realizable active colloid must operate under a finite power budget, this divergence suggests that additional physics, such as first-order reaction kinetics \citep{michelin2014phoretic, yariv2017boundary}, may be required for the model to remain relevant in sufficiently narrow gaps. 
Incorporating these effects into the present framework could reveal nontrivial behaviors of Janus particles even in the lubrication limit, clarify the validity of the model assumptions, and improve quantitative agreement with numerical and experimental studies.

\section*{Acknowledgements}We thank Ehud Yariv for helpful discussions.
\section*{Funding}

H.A.S. acknowledges support from grant no. CBET-2246791 from the US National Science Foundation. We thank the Princeton Center for Complex Materials, a MRSEC (NSF DMR-2011750), for support of this research.

\section*{Declaration of interests}The authors report no conflict of interest.
\appendix
\section{Solute concentration: transition-region analysis}\label{transition_region}
In Sec.~\ref{Axi_section}, we analyzed solute transport in the narrow gap between the particle and the wall using stretched cylindrical coordinates in (\ref{scaling_length_velocity}). We found that the active cap and the inert face correspond, at leading order in $\epsilon$, to the regions $R > \Phi$ and $0 \leq R < \Phi$, respectively. The leading-order solute concentration gradients in these two regions were obtained in (\ref{leading_gradient}). However, there remains a constant $K_+$, which needs to be determined from the matching condition at $R = \Phi$. In this appendix, we justify the claim that the transition requires the solute concentration gradient $dC^{(0)}/dR$ to be continuous at $R = \Phi$, yielding the value of $K_+$ as given in (\ref{constant_leading}). To examine the structure of the transition region near $R = \Phi$, we introduce a further stretched coordinate system centered at this interface,
\begin{equation}
\mathcal{R} = (R-\Phi)/\epsilon^{1/2} \quad \text{and} \quad \mathcal{Z} = Z.
\label{transition_variables}
\end{equation}
In the original coordinates $(r,z)$, the variables defined in~(\ref{transition_variables}) describe a region of radial and vertical extents both of $O(\epsilon)$, centered at the boundary between the active cap and the inert face. In terms of $\mathcal{R}$ and $\mathcal{Z}$, the two spatial directions are comparable in magnitude, suggesting that the leading-order problem is governed by the Laplace equation. Substituting~(\ref{transition_variables}) into~(\ref{gap_laplace_axisymmetric}) yields
\begin{equation}
\frac{\partial^2 C}{\partial \mathcal{Z}^2} + \frac{\Phi}{\Phi + \epsilon^{1/2}\mathcal{R}}\frac{\partial^2 C}{\partial \mathcal{R}^2} + 
\frac{\epsilon^{1/2}}{\Phi + \epsilon^{1/2}\mathcal{R}}\frac{\partial}{\partial \mathcal{R}}\left(\mathcal{R} \frac{\partial C}{\partial \mathcal{R}}\right) = 0.
\label{transition_region_equation}
\end{equation}
On the other hand, the solute flux boundary conditions (\ref{gap_activity}) take the form
\begin{equation}
   \frac{1}{\epsilon}\left(\frac{\partial C}{\partial \mathcal{Z}} - \Phi\epsilon^{1/2} \frac{\partial C}{\partial \mathcal{R}} + O(\epsilon)\right)\left(1 + O(\epsilon)\right) = 
    \begin{cases}
        1,\quad \, \mathcal{R} > 0 + O\left(\epsilon^{1/2}\right)\\
        0,\quad \, \mathcal{R} < 0 + O\left(\epsilon^{1/2}\right)
    \end{cases}
    \quad\text{at}\quad \mathcal{Z} = 1+\frac{\Phi^2}{2} + O\left(\epsilon^{1/2}\right).
    \label{gap_activity_transition}
\end{equation}
In the transition region, we seek a perturbation expansion for $C$ in powers of $\epsilon^{1/2}$,
\begin{equation}
C(R,Z;\epsilon) = \mathsf{C}^{(0)}(\mathcal{R},\mathcal{Z}) + \epsilon^{1/2} \mathsf{C}^{(1)}(\mathcal{R},\mathcal{Z}) + \epsilon \mathsf{C}^{(2)}(\mathcal{R},\mathcal{Z}) + ...
\label{transition_expansion}
\end{equation}
The matching conditions for the gap-scale solution $C^{(0)}$ across $R = \Phi$ depend on the behavior of $\mathsf{C}^{(0)}$ and $\mathsf{C}^{(1)}$ in the transition region. To see this, we rewrite the gap solution (\ref{leading_gradient}) in terms of the variables $\mathcal{R}$ and $\mathcal{Z}$ and expand to $O(\epsilon)$ as, 
\begin{equation}\label{matching_plus}
C_+^{(0)} = K'_ {+} -  \frac{\epsilon^{1/2} \mathcal{R}}{\Phi}\left(1 - \frac{K_{+}}{1+ \frac{\Phi^2}{2}}\right) + O(\epsilon),
\end{equation}
where $K'_ {+} $ is a constant. Note that the exact value of $K'_{+}$ is not needed to determine $K_{+}$, nor does it enter the slip velocity (\ref{Stokes_bc}). Similarly, expanding $C_-^{(0)}$ in the gap-scale region in terms of $\mathcal{R}$ and $\mathcal{Z}$ yields 
\begin{equation}\label{matching_minus}
C_-^{(0)} = K'_ {-}  + O(\epsilon),
\end{equation}
where $K'_ {-} $ is another constant. By the matching principle, \eqref{matching_plus} and \eqref{matching_minus} serve as effective boundary conditions for the transition-region solutions in the limits $\mathcal{R} \longrightarrow \pm \infty$. In particular,
\begin{equation}
\mathsf{C}^{(0)} = K'_{+} \quad \text{and} \quad \mathsf{C}^{(1)} = -\frac{\mathcal{R}}{\Phi}\left(1 - \frac{K_{+}}{1+ \frac{\Phi^2}{2}}\right)  \quad \text{as} \quad \mathcal{R} \longrightarrow +\infty,
\label{transition_bc_posinf}
\end{equation}
and
\begin{equation}
\mathsf{C}^{(0)} = K'_{-} \quad \text{and} \quad \mathsf{C}^{(1)} = 0 \quad \text{as} \quad \mathcal{R} \longrightarrow -\infty.
\label{transition_bc_neginf}
\end{equation} 

Substituting the expansion (\ref{transition_expansion}) into (\ref{transition_region_equation}), we find that the leading-order concentration $\mathsf{C}^{(0)}$ satisfies Laplace’s equation in the variables $(\mathcal{R},\mathcal{Z})$ over the strip
$(\mathcal{R},\mathcal{Z}) \in \mathbb{R} \times \left(0, 1+\frac{\Phi^2}{2}\right)$,
\begin{equation}
\nabla_{\mathcal{R},\mathcal{Z}}^2\mathsf{C}^{(0)} = \frac{\partial^2 \mathsf{C}^{(0)}}{\partial \mathcal{R}^2} + \frac{\partial^2 \mathsf{C}^{(0)}}{\partial \mathcal{Z}^2}  = 0.
\label{laplace_transition_region}
\end{equation}
This equation is supplemented with no-flux boundary conditions at the particle surface (\ref{gap_activity_transition}) and at the wall (\ref{wall_bcs}),
\begin{equation}
\frac{\partial \mathsf{C}^{(0)}}{\partial \mathcal{Z}} = 0 \quad \text{at} \quad \mathcal{Z} = 1 + \frac{\Phi^2}{2} \quad \text{and} \quad \frac{\partial \mathsf{C}^{(0)}}{\partial \mathcal{Z}} = 0 \quad \text{at} \quad \mathcal{Z} = 0.
\label{wall_bcs_transition}
\end{equation}
The solution of (\ref{laplace_transition_region}), subject to the boundary conditions (\ref{wall_bcs_transition}) and the far-field conditions (\ref{transition_bc_posinf}) and (\ref{transition_bc_neginf}), is simply a constant. Consequently, the far-field constants must coincide, i.e., $K'_{+} = K'_{-}$, and are equal to $\mathsf{C}^{(0)}$. We note that any non-constant eigenfunctions are
implicitly excluded by the matching procedure \citep{jeffrey1978temperature}. Although the analysis at this order is trivial, it implies that the leading-order excess solute concentration in the gap $C^{(0)}$ must be continuous across $R = \Phi$.

Because $\mathsf{C}^{(0)}$ is a constant, the governing equation (\ref{transition_region_equation}) and the boundary condition
(\ref{gap_activity_transition}) can be considerably simplified when expanded at $O\left (\epsilon^{1/2}\right )$. In particular, we find that $\mathsf{C}^{(1)}$ also satisfies the Laplace equation, 
\begin{equation}
\nabla^2_{\mathcal{R}, \mathcal{Z}}\mathsf{C}^{(1)} = 0, 
\label{laplace_transition_first_order}
\end{equation}
together with no-flux boundary conditions at the particle surface and at the wall, identical to those in (\ref{wall_bcs_transition}) for $\mathsf{C}^{(0)}$. This is not coincidental, as the inhomogeneous flux in (\ref{gap_activity_transition}) first appears at $O(\epsilon)$.  Note that the analysis for $\mathsf{C}^{(1)}$ differs from that for $\mathsf{C}^{(0)}$ only through the far-field boundary condition (\ref{transition_bc_posinf}). We integrate (\ref{laplace_transition_first_order}) over the domain $\left(-\tilde{\mathcal{R}}, \tilde{\mathcal{R}}\right) \times \left(0, 1 + \frac{\Phi^2}{2}\right)$, take the limit $\tilde{\mathcal{R}} \longrightarrow +\infty$, and apply the divergence theorem to conclude that the net solute flux in the far field vanishes, as a consequence of the no-flux boundary conditions on both the particle surface and the wall. This gives,
\begin{equation}
\frac{\partial \mathsf{C}^{(1)}}{\partial \mathcal{R}} \bigg|_{\mathcal{R} \longrightarrow +\infty} = -\frac{1}{\Phi}\left(1 - \frac{K_{+}}{1+ \frac{\Phi^2}{2}}\right) = 0 \quad \Longrightarrow \quad  K_{+} = 1 + \frac{\Phi^2}{2}.
\end{equation}
For this value of $K_{+}$, we find that $dC_{+}^{(0)}/dR = dC_{-}^{(0)}/dR = 0$ at $R = \Phi$. Therefore, the transition-region analysis shows that both $C^{(0)}$ and $dC^{(0)}/dR$ must be continuous across $R = \Phi$. However, the second derivative $d^2C^{(0)}/dR^2$ need not be continuous at $R = \Phi$. This lack of continuity of the second derivative  allows for
non-trivial solutions in Sec.~\ref{Tilted_section} when the Janus particle is slightly tilted, see (\ref{reduced_first_boundary_condition_tilted}). Analyzing the transition region via a matching procedure, rather than patching solutions across $R = \Phi$, thus provides precise control over the regularity of the solution at each order in the asymptotic series.

\section{Solute concentration: a connection formula across a general catalytic edge}\label{general_transition_region}
We derive the necessary conditions to be imposed on the leading-order solute concentration across a general catalytic edge. This extends the analysis in Appendix~\ref{transition_region} to domains with angular dependence and provides the closure needed to determine the solute concentration in Sec.~\ref{Solute_tilted_section}.

Let $\Omega$ denote the leading-order inert-face region in the lateral plane, expressed in polar coordinates $(R,\theta)$, with a smooth, non-self-intersecting boundary $\Gamma=\partial\Omega$. For example, in the axisymmetric case discussed in Sec.~\ref{Axi_section}, $\Omega$ is given by $R<\Phi$, whereas in the tilted case discussed in Sec.~\ref{Tilted_section}, $\Omega$ is given by $R<\Phi+\Psi\cos\theta$. The leading-order solute concentration in the gap is described by two outer solutions, one on either side of $\Gamma$. Specifically,
\begin{equation*}
    C_-^{(0)} \:\: \text{inside } \Omega,
    \qquad
    C_+^{(0)} \:\: \text{outside } \Omega .
\end{equation*}
We claim that an analysis of the local transition region near $\Gamma$ yields continuity of the leading-order solute concentration and its normal derivative across $\Gamma$:
\begin{equation}\label{connection_formula}
C_{+}^{(0)} = C_{-}^{(0)}
\quad \text{and} \quad
\tilde{\mathbf{n}}\cdot \nabla C_{+}^{(0)}
=
\tilde{\mathbf{n}}\cdot \nabla C_{-}^{(0)}
\quad \text{on} \quad  \Gamma,
\end{equation}
where $\tilde{\mathbf{n}}$ is the unit outward normal to $\Gamma$.

We choose a reference point on $\Gamma$ and let $\boldsymbol{\gamma}(s)$ be an arclength parametrization of the corresponding curve in the lateral plane, where $s$ denotes arclength measured along $\Gamma$ from the reference point. A point $\mathbf{X}$ sufficiently close to $\Gamma$ can be written as 
\begin{equation}
    \mathbf{X} = \boldsymbol{\gamma}(s) + n\tilde{\mathbf{n}}(s),
\end{equation}
where $n$ is the signed distance from the edge. With the above convention,
\begin{equation*}
    n < 0 \:\: \text{inside } \Omega,
    \qquad
    n > 0 \:\: \text{outside } \Omega.
\end{equation*}

We now introduce stretched coordinates centered on $\Gamma$, in parallel with~(\ref{transition_variables}),
\begin{equation}
    \eta = n/\epsilon^{1/2}
    \quad \text{and} \quad
    \mathcal{Z} = Z .
\end{equation}
To leading order, the local gap height is given by its value on the edge:
\begin{equation}
    H_\Gamma(s) = H(\boldsymbol{\gamma}(s))
    = 1 + \frac{1}{2} |\boldsymbol{\gamma}(s)|^2,
\end{equation}
where we have used the parabolic gap profile $H=1+R^2/2$.

Analogous to~\eqref{transition_expansion}, we seek a perturbation expansion for $C$ in powers of $\epsilon^{1/2}$ in the transition region near the boundary $\Gamma$,
\begin{equation}
C(R,\theta,Z;\epsilon)
=
\mathsf{C}^{(0)}(s,\eta,\mathcal{Z})
+\epsilon^{1/2}\mathsf{C}^{(1)}(s,\eta,\mathcal{Z})
+\epsilon \mathsf{C}^{(2)}(s,\eta,\mathcal{Z})
+ ...
\end{equation}
Recall from~\eqref{gap_laplace_tilted} that the solute concentration in the gap satisfies
\begin{equation}
\frac{\partial^2 C}{\partial Z^2}
+
\epsilon \nabla^2_{\parallel} C
=
0,
\end{equation}
where $\nabla^2_{\parallel}$ denotes the lateral Laplacian. In the orthogonal coordinates $(s,n)$, this operator is
\begin{equation}\label{lateral_laplacian}
\nabla^2_{\parallel} C
=
\frac{1}{1+\kappa n}
\left(
\frac{\partial}{\partial s}
\left(
\frac{1}{1+\kappa n}
\frac{\partial C}{\partial s}
\right)
+
\frac{\partial}{\partial n}
\left(
(1+\kappa n)
\frac{\partial C}{\partial n}
\right)
\right),
\end{equation}
where $\kappa=\kappa(s)$ is the signed curvature of $\Gamma$, which is of order $O(1)$. Since $n=\epsilon^{1/2}\eta$, it follows from~\eqref{lateral_laplacian} that, in the transition region, the lateral Laplacian admits the form
\begin{equation}
\nabla^2_{\parallel} C
=
\frac{1}{\epsilon}
\frac{\partial^2 C}{\partial \eta^2}
+
O(\epsilon^{-1/2}).
\end{equation}
We then conclude that, in the transition region, the leading-order solute concentration $\mathsf{C}^{(0)}$ satisfies Laplace's equation in the variables $(\eta,\mathcal{Z})$ over the strip
$(\eta,\mathcal{Z}) \in \mathbb{R} \times \left(0, H_\Gamma(s)\right)$,
\begin{equation}
\nabla_{\eta,\mathcal{Z}}^2\mathsf{C}^{(0)} = \frac{\partial^2 \mathsf{C}^{(0)}}{\partial \eta^2} + \frac{\partial^2 \mathsf{C}^{(0)}}{\partial \mathcal{Z}^2}  = 0.
\label{laplace_general_transition_region}
\end{equation}

Since $\partial C/\partial Z = O(\epsilon)$ by~\eqref{gap_activity_tilted}, $\mathsf{C}^{(0)}$ and $\mathsf{C}^{(1)}$ both satisfy homogeneous Neumann conditions at the wall $\mathcal{Z}=0$ and at the particle surface $\mathcal{Z}=H_{\Gamma}(s)$. The boundary conditions as $\eta \to \pm \infty$ are obtained by expanding the outer solute concentration in the gap for small $n$,
\begin{align}
    C_\pm^{(0)}(s,n)
    &= C_\pm^{(0)}(s,0)
    + n\,\frac{\partial C_\pm^{(0)}}{\partial n}(s,0)
    + O(n^2), \quad (|n| \ll 1) \\
  &= C_\pm^{(0)}(s,0)
    + \epsilon^{1/2}\eta \frac{\partial C_\pm^{(0)}}{\partial n}(s,0)
    + O(\epsilon).
\end{align}
Asymptotic matching from the transition region to the gap region requires
\begin{equation}
    \mathsf{C}^{(0)} \longrightarrow C_\pm^{(0)}(s,0) \quad \text{and} \quad \mathsf{C}^{(1)} \longrightarrow \eta \frac{\partial C_\pm^{(0)}}{\partial n}(s,0) \quad \text{as} \quad \eta \longrightarrow \pm \infty.
\end{equation}
The remainder of the argument proceeds as in Appendix~\ref{transition_region}. In particular, the leading-order problem for $\mathsf{C}^{(0)}$ admits only a constant bounded solution, which yields continuity of the leading-order solute concentration across $\Gamma$:
\begin{equation}
C_+^{(0)} = C_-^{(0)} .
\end{equation}
This constant leading-order solution simplifies the $O(\epsilon^{1/2})$ problem, from which $\mathsf{C}^{(1)}$ is again found to satisfy
\begin{equation}\label{laplace_general_transition_first}
\nabla_{\eta,\mathcal{Z}}^2 \mathsf{C}^{(1)} = 0 .
\end{equation}
Since $\mathsf{C}^{(1)}$ satisfies homogeneous Neumann conditions at $Z=0$ and $Z=H_\Gamma(s)$, we integrate~\eqref{laplace_general_transition_first} over the domain $(-\tilde{\eta},\tilde{\eta})\times(0,H_\Gamma(s))$ and apply the divergence theorem. This yields
\begin{equation}
\int_0^{H_\Gamma(s)}
\frac{\partial \mathsf C^{(1)}}{\partial \eta}(\tilde{\eta},\mathcal Z)
\,d\mathcal Z
-
\int_0^{H_\Gamma(s)}
\frac{\partial \mathsf C^{(1)}}{\partial \eta}(-\tilde{\eta},\mathcal Z)
\,d\mathcal Z
=0 .
\end{equation}
Taking the limit $\tilde{\eta}\to+\infty$ and using the far-field matching condition for $\mathsf C^{(1)}$ then gives
\begin{equation}
H_\Gamma(s)
\left(
\frac{\partial C_+^{(0)}}{\partial n}(s,0)
-
\frac{\partial C_-^{(0)}}{\partial n}(s,0)
\right)
=0 .
\end{equation}
Since $H_\Gamma(s)>0$, this implies
\begin{equation}
\frac{\partial C_+^{(0)}}{\partial n}(s,0)
=
\frac{\partial C_-^{(0)}}{\partial n}(s,0).
\end{equation}
Together, these two matching conditions yield~\eqref{connection_formula}, as claimed. 

\bibliography{literature}

@article{mozaffari2016self,
  title={Self-diffusiophoretic colloidal propulsion near a solid boundary},
  author={Mozaffari, Ali and Sharifi-Mood, Nima and Koplik, Joel and Maldarelli, Charles},
  journal={Physics of Fluids},
  volume={28},
  number={5},
  year={2016},
  publisher={AIP Publishing}
}

@article{anderson1989colloid,
  title={Colloid transport by interfacial forces},
  author={Anderson, John L},
  journal={Annual Review of Fluid Mechanics},
  volume={21},
  number={1},
  pages={61--99},
  year={1989},
  publisher={Annual Reviews 4139 El Camino Way, PO Box 10139, Palo Alto, CA 94303-0139, USA}
}

@article{golestanian2007designing,
  title={Designing phoretic micro-and nano-swimmers},
  author={Golestanian, Ramin and Liverpool, TB and Ajdari, A},
  journal={New Journal of Physics},
  volume={9},
  number={5},
  pages={126},
  year={2007},
  publisher={IOP Publishing}
}

@article{cordova2008osmotic,
  title={Osmotic propulsion: the osmotic motor},
  author={C{\'o}rdova-Figueroa, Ubaldo M and Brady, John F},
  journal={Physical Review Letters},
  volume={100},
  number={15},
  pages={158303},
  year={2008},
  publisher={APS}
}

@article{julicher2009generic,
  title={Generic theory of colloidal transport},
  author={J{\"u}licher, Frank and Prost, Jacques},
  journal={The European Physical Journal E},
  volume={29},
  number={1},
  pages={27--36},
  year={2009},
  publisher={Springer}
}

@article{sabass2012dynamics,
  title={Dynamics and efficiency of a self-propelled, diffusiophoretic swimmer},
  author={Sabass, Benedikt and Seifert, Udo},
  journal={The Journal of Chemical Physics},
  volume={136},
  number={6},
  year={2012},
  publisher={AIP Publishing}
}

@article{michelin2014phoretic,
  title={Phoretic self-propulsion at finite P{\'e}clet numbers},
  author={Michelin, S{\'e}bastien and Lauga, Eric},
  journal={Journal of Fluid Mechanics},
  volume={747},
  pages={572--604},
  year={2014},
  publisher={Cambridge University Press}
}

@article{ebbens2010pursuit,
  title={In pursuit of propulsion at the nanoscale},
  author={Ebbens, Stephen J and Howse, Jonathan R},
  journal={Soft Matter},
  volume={6},
  number={4},
  pages={726--738},
  year={2010},
  publisher={Royal Society of Chemistry}
}

@article{paxton2004catalytic,
  title={Catalytic nanomotors: autonomous movement of striped nanorods},
  author={Paxton, Walter F and Kistler, Kevin C and Olmeda, Christine C and Sen, Ayusman and St. Angelo, Sarah K and Cao, Yanyan and Mallouk, Thomas E and Lammert, Paul E and Crespi, Vincent H},
  journal={Journal of the American Chemical Society},
  volume={126},
  number={41},
  pages={13424--13431},
  year={2004},
  publisher={ACS Publications}
}

@article{howse2007self,
  title={Self-motile colloidal particles: from directed propulsion to random walk},
  author={Howse, Jonathan R and Jones, Richard AL and Ryan, Anthony J and Gough, Tim and Vafabakhsh, Reza and Golestanian, Ramin},
  journal={Physical Review Letters},
  volume={99},
  number={4},
  pages={048102},
  year={2007},
  publisher={APS}
}

@article{ebbens2011direct,
  title={Direct observation of the direction of motion for spherical catalytic swimmers},
  author={Ebbens, Stephen J and Howse, Jonathan R},
  journal={Langmuir},
  volume={27},
  number={20},
  pages={12293--12296},
  year={2011},
  publisher={ACS Publications}
}

@article{jiang2010active,
  title={Active motion of a Janus particle by self-thermophoresis in a defocused laser beam},
  author={Jiang, Hong-Ren and Yoshinaga, Natsuhiko and Sano, Masaki},
  journal={Physical Review Letters},
  volume={105},
  number={26},
  pages={268302},
  year={2010},
  publisher={APS}
}

@article{ebbens2012size,
  title={Size dependence of the propulsion velocity for catalytic Janus-sphere swimmers},
  author={Ebbens, Stephen and Tu, Mei-Hsien and Howse, Jonathan R and Golestanian, Ramin},
  journal={Physical Review E},
  volume={85},
  number={2},
  pages={020401},
  year={2012},
  publisher={APS}
}

@article{yariv2016wall,
  title={Wall-induced self-diffusiophoresis of active isotropic colloids},
  author={Yariv, Ehud},
  journal={Physical Review Fluids},
  volume={1},
  number={3},
  pages={032101},
  year={2016},
  publisher={APS}
}

@article{yariv2017boundary,
  title={Boundary-induced autophoresis of isotropic colloids: anomalous repulsion in the lubrication limit},
  author={Yariv, Ehud},
  journal={Journal of Fluid Mechanics},
  volume={812},
  pages={26--40},
  year={2017},
  publisher={Cambridge University Press}
}

@article{volpe2011microswimmers,
  title={Microswimmers in patterned environments},
  author={Volpe, Giovanni and Buttinoni, Ivo and Vogt, Dominik and K{\"u}mmerer, Hans-J{\"u}rgen and Bechinger, Clemens},
  journal={Soft Matter},
  volume={7},
  number={19},
  pages={8810--8815},
  year={2011},
  publisher={Royal Society of Chemistry}
}

@article{kreuter2013transport,
  title={Transport phenomena and dynamics of externally and self-propelled colloids in confined geometry},
  author={Kreuter, Christian and Siems, Ullrich and Nielaba, Peter and Leiderer, Paul and Erbe, Artur},
  journal={The European Physical Journal Special Topics},
  volume={222},
  number={11},
  pages={2923--2939},
  year={2013},
  publisher={Springer}
}

@article{das2015boundaries,
  title={Boundaries can steer active Janus spheres},
  author={Das, Sambeeta and Garg, Astha and Campbell, Andrew I and Howse, Jonathan and Sen, Ayusman and Velegol, Darrell and Golestanian, Ramin and Ebbens, Stephen J},
  journal={Nature Communications},
  volume={6},
  number={1},
  pages={8999},
  year={2015},
  publisher={Nature Publishing Group UK London}
}

@article{brown2016swimming,
  title={Swimming in a crystal},
  author={Brown, Aidan T and Vladescu, Ioana D and Dawson, Angela and Vissers, Teun and Schwarz-Linek, Jana and Lintuvuori, Juho S and Poon, Wilson CK},
  journal={Soft Matter},
  volume={12},
  number={1},
  pages={131--140},
  year={2016},
  publisher={Royal Society of Chemistry}
}

@article{uspal2016guiding,
  title={Guiding catalytically active particles with chemically patterned surfaces},
  author={Uspal, WE and Popescu, Mikhail N and Dietrich, Siegfried and Tasinkevych, Mykola},
  journal={Physical Review Letters},
  volume={117},
  number={4},
  pages={048002},
  year={2016},
  publisher={APS}
}

@article{simmchen2016topographical,
  title={Topographical pathways guide chemical microswimmers},
  author={Simmchen, Juliane and Katuri, Jaideep and Uspal, William E and Popescu, Mihail N and Tasinkevych, Mykola and S{\'a}nchez, Samuel},
  journal={Nature Communications},
  volume={7},
  number={1},
  pages={10598},
  year={2016},
  publisher={Nature Publishing Group UK London}
}

@article{uspal2015self,
  title={Self-propulsion of a catalytically active particle near a planar wall: from reflection to sliding and hovering},
  author={Uspal, WE and Popescu, Mikhail N and Dietrich, S and Tasinkevych, M},
  journal={Soft Matter},
  volume={11},
  number={3},
  pages={434--438},
  year={2015},
  publisher={Royal Society of Chemistry}
}

@article{ibrahim2016walls,
  title={How walls affect the dynamics of self-phoretic microswimmers},
  author={Ibrahim, Yahaya and Liverpool, Tanniemola B},
  journal={The European Physical Journal Special Topics},
  volume={225},
  number={8},
  pages={1843--1874},
  year={2016},
  publisher={Springer}
}

@article{bayati2019dynamics,
  title={Dynamics near planar walls for various model self-phoretic particles},
  author={Bayati, Parvin and Popescu, Mihail N and Uspal, William E and Dietrich, S and Najafi, Ali},
  journal={Soft Matter},
  volume={15},
  number={28},
  pages={5644--5672},
  year={2019},
  publisher={Royal Society of Chemistry}
}

@article{das2020floor,
  title={Floor-or ceiling-sliding for chemically active, gyrotactic, sedimenting Janus particles},
  author={Das, Sayan and Jalilvand, Zohreh and Popescu, Mihail N and Uspal, William E and Dietrich, Siegfried and Kretzschmar, Ilona},
  journal={Langmuir},
  volume={36},
  number={25},
  pages={7133--7147},
  year={2020},
  publisher={ACS Publications}
}

@article{ibrahim2015dynamics,
  title={The dynamics of a self-phoretic Janus swimmer near a wall},
  author={Ibrahim, Yahaya and Liverpool, Tanniemola B},
  journal={Europhysics Letters},
  volume={111},
  number={4},
  pages={48008},
  year={2015},
  publisher={IOP Publishing}
}

@article{uspal2019active,
  title={Active Janus colloids at chemically structured surfaces},
  author={Uspal, WE and Popescu, Mihail N and Dietrich, Siegfried and Tasinkevych, M},
  journal={The Journal of Chemical Physics},
  volume={150},
  number={20},
  year={2019},
  publisher={AIP Publishing}
}

@article{turk_singh_stone_2025, 
    title={Autophoretic skating along permeable surfaces}, 
    volume={1019},  
    journal={Journal of Fluid Mechanics}, 
    author={Turk, Günther and Singh, Rajesh and Stone, Howard A.}, 
    year={2025}, 
    pages={A57}
}

@article{cox1967slow,
  title={The slow motion of a sphere through a viscous fluid towards a plane surface—II Small gap widths, including inertial effects},
  author={Cox, Raymond G and Brenner, Howard},
  journal={Chemical Engineering Science},
  volume={22},
  number={12},
  pages={1753--1777},
  year={1967},
  publisher={Elsevier}
}

@article{jeffrey1978temperature,
  title={The temperature field or electric potential around two almost touching spheres},
  author={Jeffrey, DJ and Van Dyke, M},
  journal={IMA Journal of Applied Mathematics},
  volume={22},
  number={3},
  pages={337--351},
  year={1978},
  publisher={Oxford University Press}
}

@article{goldman1967slow,
  title={Slow viscous motion of a sphere parallel to a plane wall—I Motion through a quiescent fluid},
  author={Goldman, Arthur Joseph and Cox, Raymond G and Brenner, Howard},
  journal={Chemical Engineering Science},
  volume={22},
  number={4},
  pages={637--651},
  year={1967},
  publisher={Elsevier}
}

@article{ketzetzi2020slip,
  title={Slip length dependent propulsion speed of catalytic colloidal swimmers near walls},
  author={Ketzetzi, Stefania and De Graaf, Joost and Doherty, Rachel P and Kraft, Daniela J},
  journal={Physical Review Letters},
  volume={124},
  number={4},
  pages={048002},
  year={2020},
  publisher={APS}
}

@article{ketzetzi2020diffusion,
  title={Diffusion-based height analysis reveals robust microswimmer-wall separation},
  author={Ketzetzi, Stefania and De Graaf, Joost and Kraft, Daniela J},
  journal={Physical Review Letters},
  volume={125},
  number={23},
  pages={238001},
  year={2020},
  publisher={APS}
}

@article{brown2014ionic,
  title={Ionic effects in self-propelled Pt-coated Janus swimmers},
  author={Brown, Aidan and Poon, Wilson},
  journal={Soft Matter},
  volume={10},
  number={22},
  pages={4016--4027},
  year={2014},
  publisher={Royal Society of Chemistry}
}

@article{campbell2013gravitaxis,
  title={Gravitaxis in spherical Janus swimming devices},
  author={Campbell, Andrew I and Ebbens, Stephen J},
  journal={Langmuir},
  volume={29},
  number={46},
  pages={14066--14073},
  year={2013},
  publisher={ACS Publications}
}
\end{document}